\documentclass[english]{iopart}

\usepackage{graphicx}
\usepackage{amssymb}

\newcounter{fig}
\begin{document}

\title[Scaling functions]{\Large
Scaling functions in the square Ising model }

\author{ 
 S. Hassani$^\S$ and J.-M. Maillard$^{||}$}
\address{\S  Centre de Recherche Nucl\'eaire d'Alger, 
2 Bd. Frantz Fanon, BP 399, 16000 Alger, Algeria}
\address{$||$ LPTMC,  UMR 7600 CNRS, Universit\'e de Paris 6, Sorbonne Universit\'es, 
Tour 24, 4\`eme \'etage, case 121, 
 4 Place Jussieu, 75252 Paris Cedex 05, France} 

\begin{abstract}

We show and give the linear differential operators $\, {\cal L}^{scal}_q$
of order $\, q= \, n^2/4+n+7/8+(-1)^n/8$,  
for the integrals $\,I_n(r)$ which appear in the two-point correlation 
scaling function of Ising model 
$\, F_{\pm}(r) \,=\,\,  \lim_{scaling} {\cal M}_{\pm}^{-2} \, < \sigma_{0,0} \, \sigma_{M,N}>
\,=\,\,  \sum_{n} I_{n}(r)$.
The integrals $\, I_{n}(r)$ are given in expansion around $\, r= \, 0$ in the basis of the
formal solutions of $\, {\cal L}^{scal}_q$ with transcendental combination coefficients.
We find that the expression $ \, r^{1/4}\,\exp(r^2/8)$ is a solution of 
the Painlev\'e VI equation in the scaling limit. 
Combinations of the (analytic at $\, r=\, 0$) solutions of $\, {\cal L}^{scal}_q$ 
sum to $\, \exp(r^2/8)$. 
We show that the expression $ \, r^{1/4}\,\exp(r^2/8)$
is the scaling limit of the correlation function $\, C(N, N)$ 
and $\, C(N, N+1)$. The differential Galois groups 
of the factors occurring in the operators  $\, {\cal L}^{scal}_q$ are given. 

\vskip .2cm

\vskip .4cm


\vskip .1cm

\vskip .1cm


\vskip .1cm

\vskip .1cm

\noindent 
{\bf AMS Classification scheme numbers}:
 34M55, 47E05, 81Qxx, 32G34, 34Lxx, 34Mxx, 14Kxx 

\end{abstract} 

\section{Introduction}
\label{Introd}

The scaling functions of the two-point 
correlation function of the square lattice Ising 
model $\, F_{\pm}(r)$ have been obtained by Wu et al.~\cite{wu-mc-tr-ba-76}. 
These scaling functions $\, F_{\pm}(r)$ are solutions of a {\em Painlev\'e} like 
equation~\cite{wu-mc-tr-ba-76,mc-tr-wu-77}. 
Symmetrical forms of these scaling functions have been also obtained by 
Palmer and Tracy~\cite{pal-tra-81,palmer-07}
\begin{eqnarray}
\hspace{-0.95in} && \quad  \qquad  \qquad  \qquad 
 F_{\pm}(r) \,\, = \,\, \,  \sum_n\, \,  I_n, 
\end{eqnarray}
where the $\, I_n$'s are $\, n$-dimensional integrals.

The expressions of $\, I_1$ and $\, I_2$ are known in closed form.
Note that the integrand in the integrals $\,I_n$ (see (\ref{Fpm}) below) 
is not algebraic in the variables but it is {\em holonomic}. Thus the 
integrals $\, I_n$ must be solution of linear differential 
equations\footnote[2]{The integrals of a holonomic 
integrand are also holonomic.}.
These linear differential equations, annihilating the integrals $\, I_n$, 
are the main subject of this paper. 

\vskip 0.1cm

The paper is organized as follows.
Section \ref{recalls} contains recalls on the scaling function of the two-point 
correlation function $\, F_{\pm}(r)$ and its symmetrical forms.
Section \ref{formfactors} is a recall on the $\, f_N^{(n)}$, namely the 
{\em form factors} expansion of the diagonal correlation 
functions $\, C(N,N)$ on the square lattice. 
The linear differential operators $\, {\cal L}^{scal}_q$
of order $\,q=\,n^2/4\,+n+7/8\,+(-1)^n/8$, for 
these form factors at scaling have {\em no direct sum} decomposition. The 
general solutions of $\, f_N^{(1)}$, $\, f_N^{(2)}$ 
and $\, f_N^{(3)}$, at scaling, are given.
Once the observation that the scaling limits of $\, f_N^{(1)}$ and $\, f_N^{(2)}$ 
are identical to $\, I_1$ and $\, I_2$, we show in section \ref{diffeqIn},
 that the integrals $\, I_n$ are solutions of
the linear differential operators $\, {\cal L}^{scal}_q$. The proof is carried out 
by numerical methods, allowing
to write the integrals $\, I_n$ as an expansion of 
formal solutions of $\, {\cal L}^{scal}_q$, where 
the combination coefficients are {\em transcendental numbers}.
Section \ref{Painleve} deals with the 
{\em sigma form} of Painlev\'e VI equation that 
annihilates $\, C(N,N)$, as well as its scaling limit. We seek, 
and find, four solutions to the scaled Painlev\'e equation.
To each solution, we identify the corresponding solution of the $\, N$-dependent
sigma form of Painlev\'e VI equation. 
In Section \ref{Scalinglimit}, we show that $\, x^{1/4}\,\exp(x^2/32)$
is the scaling limit of $\, C(N,N)$, and, in section \ref{AppCNNp1},
we show that it is {\em also the scaling limit of} $\, C(N,N+1)$.
We show, in section \ref{decomp}, that the factors of 
the linear differential operators for the
$\, f_N^{(n)}$ (as well as the corresponding operators in the scaling limit) 
have ``special'' differential Galois groups.
Our conclusions are given in section \ref{concl}.

\vskip 0.1cm

\section{Recalls on the scaling functions of the Ising model}
\label{recalls}

The scaling functions are defined as~\cite{wu-mc-tr-ba-76} 
(where $\, \, \xi \cdot r =\, \sqrt{M^2+N^2}$)
\begin{eqnarray}
\label{firstFpm}
\hspace{-0.95in} && \quad  \qquad  \qquad 
F_{\pm}(r) \,=\,\, \,
  \lim_{scaling} {\cal M}_{\pm}^{-2} \cdot \, < \sigma_{0,0} \, \sigma_{M,N}>, 
\end{eqnarray}
with ${\cal M}_{\pm} \,=\, \, \,(1- t)^{1/8}$, where $t$ is defined in 
section \ref{formfactors}.

The scaling functions obtained in~\cite{wu-mc-tr-ba-76} are, for $\, T <\,  T_c$
\begin{eqnarray}
F_{-}(r) \,\, =\, \,\, 
\exp\Bigl(- \sum_{n=1}^\infty {1 \over \pi^{2n}} g_{2n}(r)\Bigr), 
\end{eqnarray}
with
\begin{eqnarray}
\hspace{-0.95in} &&  
 g_{2n}(r) \,=\,\,  
{(-1)^n \over n} \, \int_1^\infty \, dy_1 \, \cdots  \int_1^\infty dy_{2n} \cdot \, 
 \prod_{j=1}^{2n}\,  {\frac{\exp(-r y_j)}{(y_j^2-1)^{1/2} (y_j+y_{j+1})}} \cdot \, 
\prod_{j=1}^{n} (y_{2j}^2-1), 
 \nonumber 
\end{eqnarray}
and for $\, T > \, T_c$
\begin{eqnarray}
\hspace{-0.95in} && \quad \qquad  \qquad  \quad 
F_{+}(r) \,\,\,  =\, \, \, \,X(r) \cdot \, F_{-}(r), 
\end{eqnarray}
where
\begin{eqnarray}
\hspace{-0.95in} && \quad \qquad  \qquad  \quad 
X(r) \,\, \,=\,\, \, \, \, 
\sum_{n=0}^\infty \,  {1 \over \pi^{2n+1}} \cdot \, g_{2n+1}(r),  
\end{eqnarray}
with
\begin{eqnarray}
\hspace{-0.95in} &&  \qquad  \quad \, \, \,\,
 g_{2n+1}(r) \,\,\,   = \,\, \, \, \, 
(-1)^n  \int_1^\infty \, dy_1 \, \cdots \int_1^\infty \, dy_{2n+1}  \cdot \,
 \prod_{j=1}^{2n+1}\,  {\frac{\exp(-r y_j)}{(y_j^2-1)^{1/2}}} 
\nonumber  \\
\hspace{-0.95in} &&  \quad \qquad  \qquad  \quad  \qquad 
\times \, \prod_{j=1}^{2n} {\frac{1}{y_j+y_{j+1}}} \cdot \,
\prod_{j=1}^{n} (y_{2j}^2-1). 
\end{eqnarray}

\vskip 0.1cm

It has been shown~\cite{wu-mc-tr-ba-76, mc-tr-wu-77} that the scaling
functions $\, F_{\pm}$ are remarkably given
by {\em nonlinear equations of Painlev\'e type}:
\begin{eqnarray}
\label{FpmP}
\hspace{-0.95in} && \quad  \,  \quad 
F_{\pm} (x) \,\, \,  = \,\,\, \,   {\sinh(\psi(r)/2) \choose \cosh(\psi(r)/2) } \cdot \,
\exp {1 \over 4} \, 
\int_r^{\infty} \, (\sinh(\psi)^2\, -\, ({d\psi \over ds})^2)\cdot \,  s \, ds , 
\end{eqnarray}
where $\, \psi(r)$ verifies:
\begin{eqnarray}
\label{1overr}
\hspace{-0.95in} && \quad  \quad \quad \quad  \quad \quad \qquad 
{1 \over r} \, {d \over dr} \, (r \, {\frac{d\psi}{dr}})\, 
\, \,-{1 \over 2} \, \sinh (2\psi)
 \, \,\,=\,\,\, \, 0. 
\end{eqnarray}
Setting
\begin{eqnarray}
\label{zetar}
\hspace{-0.95in} && \quad  \qquad   \qquad  \qquad 
\zeta(r) \,\,=\,\,\, r \, {d \over dr} \ln(F_{\pm}), 
\end{eqnarray}
the equation (\ref{FpmP}) becomes:
\begin{eqnarray}
\label{P5barry}
\hspace{-0.95in} && \quad \quad  \quad   \quad \,\,
(r \, \zeta '')^2 \,\,\, = \,\, \,\, \, 
4 \cdot \, (r \, \zeta '\, -\zeta)^2 \, \, 
\, - 4 \cdot \, (\zeta ')^2 \cdot \, \, (r\,  \zeta ' \, -\zeta)
\, \, \, + \, (\zeta ')^2. 
\end{eqnarray}

\vskip 0.1cm

The scaling functions are also given in a symmetrical form in~\cite{pal-tra-81}
(see also~\cite{palmer-07}).
\begin{eqnarray}
\label{Fpm}
\hspace{-0.95in} && \qquad \quad \quad \, \,\,\,
F_{+}(r) \,=\,\, \,  \sum_{n=0} \, I_{2n+1}(r), \qquad \quad 
F_{-}(r) \,=\,\,\,   1\, \,\,  + \sum_{n=1} I_{2n}(r), 
\end{eqnarray}
\begin{eqnarray}
\label{Inintegrals}
\hspace{-0.95in} &&  
I_n  \, \, = \,\, \,  {1 \over n!}\,  \int_0^\infty \, {du_1 \over 2\pi} 
\, \, \cdots \, \int_0^\infty \, {du_n \over 2\pi} \,\, 
 \prod_{i<j} \,  {\frac{(u_i-u_j)^2}{(u_i+u_j)^2}} \, 
\prod_{i=1}^n \,  {1 \over u_i} \exp( -{r \over 2} (u_i+1/u_i)).
\end{eqnarray}

A direct computation gives
\begin{eqnarray}
\label{integralI1}
\hspace{-0.95in} && \quad \quad \quad \quad \qquad \quad  \qquad 
I_1 \, \,=\,\, \, \,  {1 \over \pi} \cdot \, K_0(r).
\end{eqnarray}

A "less" direct computation yields
\begin{eqnarray}
\label{integralI2}
\hspace{-0.95in} && \quad  
I_2 \,\, \,  \, =\, \, \, \, \, 
{1 \over \pi^2} \cdot \, \Bigl(({1 \over 2}-r^2) \cdot \, K_0(r)^2
\, -r \cdot \,  K_0(r)
 \cdot \, K_1(r)\, \,  +\, r^2 \cdot \, K_1(r)^2  \Bigr), 
\end{eqnarray}
where $\, K_0$ (resp. $\, K_1$) is the (resp. derivative of the) 
{\em modified Bessel function} (see below).

\section{The linear ODE of the form factors and their scaling limit}
\label{formfactors}

The diagonal correlation functions $\, C(N, N)$ of the square Ising model 
have a form factor expansion~\cite{2007-holonomy}
\begin{eqnarray}
\label{CNdef}
\hspace{-0.95in} && \quad \quad \quad \quad \quad \,\,
C(N, N) \,\,  = \,\, \,  
(1-t)^{1/4} \cdot \, \Bigl(1 \, \, + \sum_{n=1}^\infty \, f^{(2n)}_N \Bigl),  
\qquad \quad T < T_c  
\end{eqnarray}
with  $\, t\,\,=\,\, \, \Bigl(\sinh(2E^v/k_BT)\,  \sinh(2E^h/k_BT) \Bigr)^{-2}$,  and
\begin{eqnarray}
\label{CNdefplus}
\hspace{-0.95in} && \quad \quad \quad \quad \quad \,\,
C (N, N)   \,\,  =\,\, \,  
(1-t)^{1/4} \cdot \,  \sum_{n=0}^\infty f^{(2n+1)}_N, 
 \quad \qquad T > T_c  
\end{eqnarray}
with $\,  t\, \,=\,\,\, \Bigl( (\sinh(2E^v/k_BT)\,\sinh(2E^h/k_BT) \Bigr)^2$, 
where $\, E^h$ and  $\, E^v = \,E^h$ are the horizontal and vertical interaction 
energies of the Ising model.

\vskip 0.2cm

The diagonal correlation functions $\, C(N, \,N)$ can be calculated from Toeplitz 
determinants~\cite{kaufman-onsager-49,montroll-potts-ward-63,mccoy-wu-73}. They 
are also solutions of Painlev\'e VI {\em in its sigma form}~\cite{jim-miw-80}.
The diagonal correlation functions $\,C(N,\, N)$, 
as well as the form factors $\, f^{(n)}_N$, write as {\em polynomials} 
in the complete elliptic integrals (see \ref{CnnETfjn} for some recalls).

The diagonal form factors $\, f^{(n)}_N$ are $\, n$-dimensional 
integrals~\cite{2007-holonomy} and
are annihilated by linear ODEs whose corresponding linear 
differential operators factorize,
with factors such that the $\, f^{(n)}_N$ 
are ``embedded'' in the form factors $\, f^{(n+2k)}_N$
\begin{eqnarray}
\label{prodbull}
\hspace{-0.95in} && \quad 
 \qquad \left( \prod_{k=1}^n \bullet L_{2k} \right) \, (f^{(2n-1)}_N) \,\, =\,\,\,  0, 
\quad \quad \quad  \,\,
 \left( \prod_{k=0}^n \bullet L_{2k+1} \right)  \, (f^{(2n)}_N) \,\, =\,\, \, 0, 
\end{eqnarray}
which means, for instance, that $\, f^{(1)}_N$ and $\, f^{(3)}_N$ are solutions of
the linear ODEs:
\begin{eqnarray}
\label{L2f1N}
\hspace{-0.95in} && \quad \quad \quad \quad \quad 
L_2 \,f^{(1)}_N \,\,=\,\,\, 0, \qquad \qquad L_4 \cdot L_2 \,f^{(3)}_N \,\,=\,\,\, 0.
\end{eqnarray}

The expressions of these order-$n$ linear differential operators $\, L_n$ 
have been obtained~\cite{2007-holonomy}
for generic values of $ \, N$ (they are given up to $ \, n= \, 10$ 
in~\cite{2007-holonomy}). This way, the scaling limit of these linear 
differential operators has been possible.
The scaling limit amounts to taking both the limits $ \, t  \, \rightarrow \, 1$ 
and $\, N \,\rightarrow \,\infty$ in the linear 
differential operators. This is performed with
the change of variable $\, x = \,(1-t) \cdot \, N$,
 keeping the leading order of $\,N$.

In the scaling limit, the  linear differential operators $\,L_n$ in the variable $\,t$
become linear differential operators $\, L_n^{scal}$ in the scaling variable $\,x$,
and we have shown~\cite{2007-holonomy} that the factors $\, L_n^{scal}$
solve as polynomial expressions of {\em modified
Bessel functions} of homogeneous degree.
For some purposes in the sequel and 
easy references, we recall the factors $\, L_1^{scal}$, $\, L_2^{scal}$, 
$\, L_3^{scal}$, $\, L_4^{scal}$, $\, L_5^{scal}$
 and $\, L_6^{scal}$ in \ref{Lnscal}.

\vskip 0.1cm

Call $\,\, B_0(x/2)\,$ and $\,\, K_0(x/2)\,$ the (respectively analytical at $ \, x=\, 0$, 
and logarithmic) solutions of the {\em modified Bessel} differential operator
(with $\,D_x$ the derivative\footnote[5]{Similarly, we will also use,
in this paper, the notations $\,D_t$ for $\, d/dt$ and $\,D_s$ for $\, d/ds$.} $\, d/dx$):
\begin{eqnarray}
\label{Dx2}
\hspace{-0.95in} && \quad  \qquad \qquad \quad  \quad
D_x^2\, \, \,   + {1 \over x} \cdot \, D_x \, \, \,  -{1 \over 4}. 
\end{eqnarray}
We call $\, B_1(x/2)$ and $\, K_1(x/2)$ the first derivative of, respectively,
$\, 2\,B_0(x/2)$ and $\, -2\,K_0(x/2)$.

\vskip 0.1cm

Consider the linear differential operator $\, L_4 \cdot L_2$ that annihilates the
form factors $\, f^{(1)}(N)$ and $\, f^{(3)}(N)$, and denote by
$\, L_4^{scal} \cdot L_2^{scal}$ the corresponding linear differential operators
in the scaling limit.

The general solution of $\, L_2^{scal}$ reads (omitting the argument $\, x/2$)
\begin{eqnarray}
\label{solL2scal}
\hspace{-0.95in} && \quad \quad  \quad  \qquad 
sol(L_2^{scal})  \, \, =  \, \,\, \, \,  \, 
c_1 \cdot \, B_0 \, \,\,   + c_2 \cdot \, K_0. 
\end{eqnarray}
The general solution of $\, L_4^{scal}$ reads
\begin{eqnarray}
\label{solL4sc}
\hspace{-0.95in} &&
sol(L_4^{scal})\, \,  = \, \, \, \, \, \,
c_3 \cdot \, 
\Bigl(B_0^3 \, -x \cdot \,  B_0^2 \cdot  \,B_1\,  + B_0 \cdot \, B_1^2 \, + x \cdot \, B_1^3 \Bigr)
\nonumber \\
\hspace{-0.95in} && \quad \quad  
\, + \, c_4 \cdot \, 
\Bigl( K_0^3 \, + x \cdot \, K_0^2 \cdot \,K_1 \, 
+ K_0 \cdot \, K_1^2 \, -x \cdot \,K_1^3 \Bigr)
\nonumber \\
\hspace{-0.95in} && \quad \quad  
 \, + \, c_5 \cdot \, \Bigl( B_0^2 \cdot \, (3 K_0 \, +x\,  K_1)\, 
+ \, B_1^2 \cdot \,  (K_0\, -3\, x\, K_1) \, 
  -2 \, B_0 \,B_1 \cdot \, (x\, K_0\, +K_1)
 \Bigr) 
\nonumber \\
\hspace{-0.95in} && \quad \quad 
\, + \,  c_6 \cdot \, \Bigl(K_0^2 \cdot \, (3 B_0 -x\,  B_1)\, 
+K_1^2 \cdot \, (B_0\, +3\, x \cdot \, B_1) \,
 -2\, K_0 K_1 \cdot \, (B_1 \, -x\, B_0) \Bigr), 
\nonumber 
\end{eqnarray}
and $\,\, L_4^{scal} \cdot L_2^{scal}\,$ solves as
\begin{eqnarray}
\label{solL4scale}
\hspace{-0.95in} && \quad 
 \quad sol(L_4^{scal} \cdot L_2^{scal}) \, \, \, \, =\, \,  \,  \, \,
sol(L_2^{scal})\,
 \nonumber \\
\hspace{-0.95in} && \quad \quad  \quad  \quad \quad \quad  \quad 
 - B_0 \cdot \, \int K_0 \cdot \,sol(L_4^{scal}) \,\cdot \, x \,  dx 
\, \,  \,\, 
+\, K_0 \cdot \, \int B_0 \cdot \,sol(L_4^{scal})\, \cdot \,x \, dx,
 \nonumber
\end{eqnarray}
i.e. the scaling limit of $\, f^{(1)}(N)\, +f^{(3)}(N)$ is 
{\em not a} 
{\em polynomial expression of modified Bessel functions}.

\vskip 0.1cm

Similarly, for $\, T <\,  T_c$, consider the linear differential 
operator $\, L_3 \cdot L_1$,
with the constant and $\, f^{(2)}(N)$ as solutions.
In the scaling limit the operator $\, L_3 \cdot L_1$ becomes
$\, L_3^{scal} \cdot\,  L_1^{scal}$ (with $\, L_1^{scal} = \, D_x$),
and its general solution reads:
\begin{eqnarray}
\label{solL3Dxscal}
\hspace{-0.95in} && \quad \, 
sol(L_3^{scal} \cdot D_x) \, \, \,  = \,  \,  \,    \,   \, c_0 \,
\,  \, \, \, +\, c_1 \cdot \, \Bigl((2\, -x^2) \cdot \,  B_0^2 \, 
+2\,  x \cdot \,  B_0 \cdot \,B_1\,  \,   + x^2 \cdot \, B_1^2 \Bigr)
 \nonumber  \\
\hspace{-0.95in} && \quad \quad \quad \, 
 + \, c_2 \cdot \, 
\Bigl((2  \, -x^2) \cdot \,  K_0^2\,\,  
-2\,  x \cdot \,  K_0 \cdot \,K_1  \, \, + x^2 \, K_1^2 \Bigr)
\\
\hspace{-0.95in} && \quad \quad \quad \,
\, + \,   c_3  \cdot \, \Bigl((x \cdot \, B_0 \cdot \, K_1 \, \,
+ \,x \cdot \, B_1 \cdot \, K_0\,\, 
-x^2  \cdot \,B_1 \cdot \,K_1 \,\, -(x^2 \, -2) \cdot \, B_0 \cdot \,K_0 \Bigr).
 \nonumber 
\end{eqnarray}
Note that $\,\,L_3^{scal} \cdot D_x\, $ {\em has a direct sum decomposition}
(see \ref{Lnscal}), {\em but the operators 
(in the scaling limit) of higher order have not}.

\section{Linear differential equations of the $\, I_n$ integrals (\ref{Inintegrals})}
\label{diffeqIn}

If we compare $\,I_1$ given in (\ref{integralI1}) with (\ref{solL2scal}),
 and $\,I_2$ given in (\ref{integralI2})
with (\ref{solL3Dxscal}), one remarks that the integrals are,
respectively, solution of the linear differential operator
 $\,L_2^{scal}$ and $\,L_3^{scal} \cdot D_x$, once
the correspondence $\,r \,\rightarrow\, x/2$ has been made.

\vskip 0.1cm

We now argue that the integrals $\, I_n(x)$ are solutions 
of the linear differential operator
\begin{eqnarray}
\label{Lnscalodd}
\hspace{-0.95in} && \quad \quad  \quad  \quad \, \, 
{\cal L}^{scal}_q \,\, =\, \, \,\,
 L_{n+1}^{scal} \cdot L_{n-1}^{scal} \,\cdots \, L_2^{scal},
 \qquad \, \, q \,=\,(n+2)^2/4,
\end{eqnarray}
for $ \, n$ odd, and 
\begin{eqnarray}
\label{Lnscaleven}
\hspace{-0.95in} && \quad \quad  \quad  \quad \, \, 
{\cal L}^{scal}_q= \,\, L_{n+1}^{scal} \cdot L_{n-1}^{scal} \, \cdots \, L_1^{scal}, 
\qquad \,\, q \,=\, (n+1)(n+3)/4,
\end{eqnarray}
for $\,n$ even.

This will be proved numerically for the first $\,I_n(x)$, i.e. we show that:
\begin{eqnarray}
\hspace{-0.95in} && \quad \quad
 L_2^{scal} \cdot I_1(x) \,=\,0, 
\qquad \qquad \quad  \, \, 
(L_3^{scal} \cdot L_1^{scal}) \cdot I_2(x) \,=\,0,  
\nonumber \\
\hspace{-0.95in} &&   \quad\quad
(L_4^{scal} \cdot L_2^{scal})  \cdot I_3(x) \,=\,0, \qquad 
(L_5^{scal} \cdot L_3^{scal} \cdot L_1^{scal}) \cdot I_4(x) \,=\,0.   
\end{eqnarray}

\vskip 0.1cm

Let us show the method for the integrals $\,I_1(x)$ and $\,I_2(x)$ which are 
known in closed form expressions. 

\subsection{The integrals $I_1(x)$ and $I_2(x)$}
\label{I1andI2}

With the formal solutions  of $\, L_2^{scal}$ at $\, x=\, 0$ 
\begin{eqnarray}
\label{S11}
\hspace{-0.95in} && \, \, \, 
S_1^{(1)} \, \, = \,\, \, 
S_2^{(1)} \cdot \, \ln(x) \, \,  
- \, \Bigl({x^2 \over 16} \, + {3 x^4 \over 2048} \,  + \cdots \, \Bigr), 
\quad \, \,  \, \, \, 
  S_2^{(1)} \,\, = \, \,\, 
1\,\, \, \,   +{x^2 \over 16}\, + {x^4 \over 1024} \, +  \,\cdots,
\nonumber
\end{eqnarray}
we form the generic combination
 $\,c_1^{(1)}\, S_1^{(1)} \,+ c_2^{(1)} \, S_2^{(1)}$ 
that we evaluate numerically 
(and its first derivative) at a fixed value of $\,x=\,x_0$. 
The integral $\,I_1(x)$ (and its first derivative) 
are performed numerically for the same value 
of\footnote[2]{One may also, obviously, 
compute the combination of solutions 
and $\,I_1(x)$, i.e. (\ref{S11}), at two values of $\,x$} $\,x$. 
Solving the system
\begin{eqnarray}
\label{I1vert}
\hspace{-0.95in} && \quad \quad \quad \quad \quad 
(c_1^{(1)} \cdot \, S_1^{(1)} \, + c_2^{(1)} \cdot \, S_2^{(1)}) \vert_{x=x_0} 
 \,\,\,  = \,\,\, \, I_1(x) \vert_{x=x_0},   \nonumber \\
\hspace{-0.95in} && \quad \quad \quad \quad \quad 
{\frac{d}{dx}}\, 
\Bigl(c_1^{(1)} \cdot\, S_1^{(1)} \,+ c_2^{(1)} \cdot\, S_2^{(1)}\Bigr) \vert_{x=x_0} 
 \, \,\,  = \, \,\, \,\,  
{\frac{d}{dx}} I_1(x) \vert_{x=x_0}, 
\end{eqnarray}
in the constants $\,c_1^{(1)}$ and $\,c_2^{(1)}$, one obtains
\begin{eqnarray}
\label{c1c2}
\hspace{-0.95in} && \quad  \quad  \quad \quad \quad \quad 
c_1^{(1)}  \,\,=\,\,  \,-0.31830, \qquad  \qquad 
c_2^{(1)} \, \,=\, \, \,0.25753, 
\end{eqnarray}
which are easy to recognize, since $\,I_1(x)$ is known 
(and given in (\ref{integralI1}) with $\,r=\,x/2$), as
\begin{eqnarray}
\label{c11c21}
\hspace{-0.95in} && \quad  \quad  \quad \quad \quad \quad 
c_1^{(1)} \,\,=\,\,\, -{1 \over \pi}, \qquad  \qquad 
c_2^{(1)}\, \,=\, \,\, {1 \over \pi} \cdot  \, (2\ln(2)\, -\gamma). 
   \nonumber 
\end{eqnarray}
where $\, \gamma$ is Euler's constant.

The same calculations are performed for  $ \, L_3^{scal} \cdot \,  D_x$ 
with the formal solutions written as
\begin{eqnarray}
\label{S12}
\hspace{-0.95in} && \quad \quad 
S_1^{(2)} \,\,\,  = \, \,\,\, \,  
 S_3^{(2)} \cdot \, \ln(x)^2 
\,\, \, + \,\, \Bigl({5 x^2 \over 8} + {9 x^4 \over 1024}\, 
+ {\frac {29}{221184}}\, x^6 \,+ \cdots \, \Bigr) \cdot  \, \ln(x)
\nonumber \\
\hspace{-0.95in} && \quad  \quad  \quad  \quad  \quad  \quad  \quad 
-\, \Bigl({3 x^2 \over 4} \, \,   + {x^4 \over 128} \, \, 
 +{\frac {19}{147456}} \cdot \,  x^6
\, \,  + \cdots \, \Bigr),  
 \nonumber \\
\hspace{-0.95in} && \quad \quad 
S_2^{(2)} \,\,\,  = \, \, \,\,\,   
S_3^{(2)} \cdot \, \ln(x) 
 \,\,\,   + \, ({5 x^2 \over 16}\,
 + {9 x^4 \over 2048}\, +{\frac {29}{442368}}\,{x}^{6}
 + \cdots \, ), \\
\hspace{-0.95in} && \quad \quad 
S_3^{(2)}\, \,\, =\,\,\,  \,\,  
1 \, \,\, \,      -{x^2 \over 8} \,\,   -{x^4 \over 512}\,\, 
   -{x^6 \over 36864}\, \, \,    + \, \cdots,  
\qquad \qquad  
S_4^{(2)} \,\, = \, \,\,  1.  
\nonumber
\end{eqnarray}
Similarly, the combination 
$\,c_1^{(2)} \cdot   \, S_1^{(2)} \, + c_2^{(2)} \cdot  \, S_2^{(2)} 
\, + c_3^{(2)}\cdot   \, S_3^{(2)} \, + c_4^{(2)} \cdot  \, S_4^{(2)}$, 
and its first three
derivatives are evaluated numerically at a fixed value of $\,x =\, x_0$,
 and matched to the integral $\,I_2(x)$ (and its
first three derivatives) performed numerically.
Solving in the constants $ \,c_j^{(2)}$, one obtains:
\begin{eqnarray}
\label{c12c42}
\hspace{-0.95in} &&  
 \,\, c_1^{(2)} \,\,=\,\,\, 0.0506605, \,\,\,
\quad c_2^{(2)}\, \,=\,\,\, 0.0193443, 
 \quad \,\,\,c_3^{(2)} \, \,=\,\, \,0.052507, \,\,\, 
\quad c_4^{(2)} \,\,=\, \,\, 10^{-8}.
   \nonumber 
\end{eqnarray}

Here also, since $I_2(x)$ is known (and given in (\ref{integralI2}) with $\,r= \,x/2$), 
the constants are easy to recognize
\begin{eqnarray}
\label{c12c32}
\hspace{-0.95in} && \quad \quad \quad 
c_1^{(2)} \,\, = \, \, \,{1 \over 2 \pi^2}, \, \qquad \qquad
 c_2^{(2)}\, \,=\, \,\,
{1 \over \pi^2}  \cdot \,  (1\, -2\ln(2)\, +\gamma),
 \\
\hspace{-0.95in} && \quad \quad \quad 
c_3^{(2)} \, \, = \, \,\,\,
 {1 \over 2 \pi^2} \cdot \, (1\,-2\ln(2)\, +\gamma)^2\,\,\, + {1 \over 2\pi^2},
 \qquad \qquad c_4^{(2)} \,\,=\, \,\, 0.  
  \nonumber 
\end{eqnarray}

\subsection{The integrals $\,I_3(x)$ and $\,I_4(x)$}
\label{I3I4}

Now, we consider the integral $\,I_3$ which should be a solution of 
$\, L_{4}^{scal} \cdot \, L_{2}^{scal}$, whose
local exponents at $\,x=\,0$ are $\,0,\, 0,\, 0,\, 0,\, 2,\, 2$ 
(that we note $\,0^4,\, 2^2$).
The formal solutions are written as:
\begin{eqnarray}
\label{S13}
\hspace{-0.95in} && \, 
 \, S_1^{(3)} \,=\,\, \, \,   S_4^{(3)}  \cdot \, \ln(x)^3 \,\,  \,  
+  \, \Bigl(3 \, - {21 x^2 \over 8} \, - {87 x^4 \over 2048}\,
 + \cdots \, \Bigr) \cdot \, \ln(x)^2
 \nonumber \\
\hspace{-0.95in} && \, \quad \quad \quad \quad 
+  \, \Bigl({9 \over 2}\,  + {9 x^2 \over 128}\, 
 + { 81 x^4 \over 8192}\,  +\,  \cdots \,\Bigr) \cdot \,  \ln(x)
\quad + \, \Bigl(3\, + {3 x^2 \over 64} \,
 - {75 x^4 \over 2048} \, + \, \cdots \, \Bigr),   
\nonumber \\
\hspace{-0.95in} && \, 
 \, S_2^{(3)} \,=\,\, S_4^{(3)} \cdot \, \ln(x)^2 \,
+ \, \Bigl( 2+{x^2 \over 32} \, - {x^4 \over 2048}
 \, + \cdots\, \Bigr) \cdot \, \ln(x)
\, \, +\, \Bigl({3 \over 2}\, +{3 x^2 \over 128} \, 
+ {27 x^4 \over 8192}\, + \cdots \, \Bigr), 
 \nonumber \\
\hspace{-0.95in} && \, 
\, S_3^{(3)} \,=\, \,   S_4^{(3)} \cdot \, \ln(x)
\,\,   + \, \Bigl(1 \,+{x^2 \over 64}\, 
- {x^4 \over 4096}\,   + \cdots \,\Bigr), \quad \,\,  
 S_4^{(3)} \, \,=\, \, \, 
 1\,+ {7 x^2 \over 16}\, + {7 x^4 \over 1024}\,\, + \cdots, 
  \nonumber \\
\hspace{-0.95in} && \, 
\, S_5^{(3)} \,=\, \,   S_1^{(1)}, 
\qquad  \qquad 
 S_6^{(3)} \,=\, \,   S_2^{(1)}.  
\end{eqnarray}

Similar calculations are performed, namely evaluating numerically the linear combination 
$\,  \sum_j c_j^{(3)}\, S_j^{(3)}$ 
(and its five derivatives) matching with the integral $\,  I_3(x)$ 
(and its five derivatives) at a given value of $\,  x=\,  x_0$.
Solving in the constants $\,  c_j^{(3)}$, one obtains:
\begin{eqnarray}
\label{c13c63}
\hspace{-0.95in} && \quad
  c_1^{(3)} \,=\, \, -0.0322515/3!, \quad \, \, \,   \, \, c_2^{(3)} \,=\,\,  -0.0184725/3!, 
 \quad \, \,\,  \,\,    c_3^{(3)} \,=\,-0.5789545/3!, 
\nonumber \\
\hspace{-0.95in} && \quad
  c_4^{(3)} \,=\, 0.65939377/3!, \quad \,\, \,\,  \, \,   \,  c_5^{(3)} \,=\, 0.49900435/3!,
 \quad \, \,\, \,  \, \,\,    \, c_6^{(3)} \,=\, -0.1942198/3!   
 \nonumber 
\end{eqnarray}

The constant $\,  c_1^{(3)}$ is easily recognized as $\,  -{1 \over 6 \pi^3}$ and we may guess the
constant $\,  c_2^{(3)}$ as $\,  (-1\,  +2 \ln(2)\,   - \gamma)/(2 \pi^3)$, but we have not attempted
to recognize the other constants, because the number of correct digits is rather low.
Note however, that if we evaluate, again, $\, \, I_3(x)\,   - \sum_j\,   c_j^{(3)}\, S_j^{(3)}$
with the obtained constants $\,  c_j^{(3)}$ and for other values of $\,  x_0$, 
one obtains zero with the working accuracy.

Similar calculations are done for $\,  I_4(x)$ with the basis 
of solutions at $\,  x=\,  0$ of 
$\,  L_{5}^{scal} \cdot L_{3}^{scal} \cdot L_{1}^{scal}$ 
(whose local exponents at $\,  x=\,  0$ are $\,  0^5, \,  2^3, 6$):
\begin{eqnarray}
\label{S14}
\hspace{-0.95in} &&
 \quad  S_1^{(4)} \,   \, = \,\, \,    \, \,  
  S_5^{(4)} \cdot  \,   \ln(x)^{4}\,  \, \,    
+ \, ({\frac {20}{3}}\,  +{\frac {143}{12}}\,{x}^{2}\,
  +{\frac {283}{1536}}\,{x}^{4}
+{\frac {5}{1024}}\,{x}^{6}\,\,   +\cdots \, ) \cdot \, \ln(x)^{3} 
\nonumber \\
\hspace{-0.95in} && \quad  \,
\quad  \qquad \,  + \, \Bigl({\frac {64}{3}} \, 
+{\frac {1765}{192}}\cdot \,{x}^{2} \,  
+{\frac {1933}{12288}}\cdot \,{x}^{4} \, 
-{\frac {275}{294912}}\cdot \,{x}^{6} \,\,  +\cdots \, \Bigr) \cdot \, \ln(x)^{2} 
\nonumber \\
\hspace{-0.95in} && \quad  \,
 \qquad \quad \,  
+ \, \Bigl({\frac {334}{9}}\,  +{\frac {5771}{1152}} \cdot \,{x}^{2}\,  
-{\frac {3829}{73728}}\cdot \, {x}^{4}\,  -{\frac {6509}{884736}}\cdot \,{x}^{6}
\,  +\,  \cdots\, \Bigr) \cdot \,  \ln(x)
\nonumber  \\
\hspace{-0.95in} && \quad  \,
 \qquad \quad \,  +{\frac {1549}{54}}\,  +{\frac {21505}{13824}}\cdot \,{x}^{2}\, 
 +\,  {\frac {466273}{884736}} \cdot \,{x}^{4}\, 
 +{\frac {102762373}{12740198400}} \cdot \,{x}^{6}
\, \,  \,   +\,  \cdots,   
\end{eqnarray}
\begin{eqnarray}
\hspace{-0.95in} &&  
 \quad  S_2^{(4)}\,  \,=\, \,\,  \,   S_5^{(4)} \cdot \, \ln(x)^{3} \, 
+ \, \Bigl(5 \, -{\frac {223}{256}}\cdot \,{x}^{2} \, 
-{\frac {247}{16384}}\cdot \,{x}^{4}\,  
+\,  {\frac {15}{131072}} \cdot \,{x}^{6}\, \,  
+\,  \cdots \, \Bigr) \cdot \,   \ln(x)^{2}
 \nonumber \\
\hspace{-0.95in} && \quad  \quad 
 \qquad \quad  +\,   \Bigl({\frac {32}{3}}\, \,
 -{\frac {473}{1536}}\cdot \,{x}^{2}\, \, 
+{\frac {199}{98304}} \cdot \,{x}^{4}\,
+{\frac {659}{1179648}}\cdot \,{x}^{6} \, \,
+\cdots \, \Bigr) \cdot \, \ln(x) 
\nonumber \\
\hspace{-0.95in} && \quad \quad 
 \qquad \quad +{\frac {167}{18}} \,\, +{\frac {485}{18432}}\cdot \,{x}^{2}\, \, 
-{\frac {37915}{1179648}}\cdot \,{x}^{4}\,  \,
-{\frac {8508439}{16986931200}}\cdot \,{x}^{6}\,\, \, +\, \cdots,  
\nonumber 
\end{eqnarray}
\begin{eqnarray}
\hspace{-0.95in} &&  
 \quad  S_3^{(4)} \,\,=\,\, \,S_5^{(4)}   \cdot \, \ln(x)^{2} \, \,
+ \, \Bigl({{10} \over {3}} \, -{\frac {223}{384}}\cdot \,{x}^{2}  \, 
-{\frac {247}{24576}}\cdot \,{x}^{4} \, 
+{\frac {5}{65536}}\cdot \,{x}^{6} \, +\cdots \, \Bigr) \cdot \, \ln(x) 
\nonumber \\
\hspace{-0.95in} && \quad  \quad  \qquad \quad 
+{\frac {32}{9}} \, -{\frac {473}{4608}}\,{x}^{2}\,
  +{\frac {199}{294912}}\cdot \,{x}^{4}\, 
 +{\frac {103027}{4246732800}}\cdot \,{x}^{6} \, \,\, \, \, 
+ \, \cdots,  
\nonumber \\
\hspace{-0.95in} && \quad 
 S_4^{(4)} \,\,=\,\, \,   S_5^{(4)} \cdot \,  \ln(x)\, \, \,  \, 
+\, {\frac {5}{3}} \, \, 
 -{\frac {223}{768}}\cdot \,{x}^{2}\,  \, 
-{\frac {247}{49152}}\cdot \,{x}^{4} \,  \,
-{\frac {30931}{707788800}}\cdot \,{x}^{6} \,\, \, +\, \cdots, 
\nonumber \\
\hspace{-0.95in} && \quad 
  S_5^{(4)}\, \,=\, \, \,\, \, 1\,  \,  \,  -{5 x^2 \over 4}\,  \, 
 -{\frac {5 x^4}{256}}\,   \, 
-{\frac {x^6}{2304}} \,\, \,  + \,   \cdots, 
 \nonumber \\
\hspace{-0.95in} &&  
\quad S_6^{(4)} \,=\,\,   S_1^{(2)}, \qquad \, \,\,
 S_7^{(4)} \,=\, \,  S_2^{(2)}, \qquad  \, \,\,
S_8^{(4)} \,=\, S_3^{(2)}, \qquad  \, \,\,
S_9^{(4)} \,=\,\,   S_4^{(2)}. 
\nonumber
\end{eqnarray}
The coefficients combination read
\begin{eqnarray}
\label{c14c94}
\hspace{-0.95in} && \quad 
\quad c_1^{(4)} \,=\,\,  0.0102659/4!, \quad \,\, \,\,\, \,\,   c_2^{(4)} \,=\,\,  0.0215279/4!,  
\quad \,\, \,\,\,\, \,  c_3^{(4)} \,=\,\,  0.423376/4!, 
\nonumber \\
\hspace{-0.95in} && \quad 
 \quad c_4^{(4)} \,=\,\,  -1.086613/4!, \quad \,\,\,  \,\,\,  c_5^{(4)} \,=\,\,  1.063659/4!, 
\quad \,\, \,\,\,\,\,\,\,   c_6^{(4)} \,=\,\,   - 0.35704/4!,  
\nonumber \\
\hspace{-0.95in} && \quad 
\quad c_7^{(4)} \,=\,- 0.02156/4!, \quad \quad\,\,   \,\,c_8^{(4)} \,=\,1.05496/4!, 
\quad \,\,  \,\,\, \,\,\, \, \,  \,\,\, c_9^{(4)} \,=\,- 1.38534/4!   
\nonumber 
\end{eqnarray}

Here also, the same numeric values of $\,  c_j^{(4)}$ 
are obtained for {\em any other value} of $\,  x_0$.

\vskip 0.1cm

Let us remark that if one just  wants to check that $\,  I_n(x)$ is a solution 
of $\,  {\cal L}^{scal}_q$, one may proceed as follows. Call $\,  {\cal I}_n(x, u)$ 
the integrand of $\,  I_n(x)$ and integrate numerically 
\begin{eqnarray}
\label{calLn}
\hspace{-0.95in} && \quad \quad \quad \quad \quad \quad  \quad  \quad 
{\cal L}^{scal}_q({\cal I}_n(x, u)), 
\end{eqnarray}
for fixed $\,  n$ and various values of $\,  x$, 
to get zero with the desired accuracy.

\vskip 0.1cm
 
We claim that this continues for the higher $\,  I_n$,
and conclude that the integrals $\,  I_n(x)$ are solutions 
of $\,  {\cal L}^{scal}_q$, the scaling limit 
of the linear differential operator annihilating
 the form factors $\,  f_N^{(n)}$. 

\subsection{The expansion around $\, x=\, 0$ of the integrals $\, I_n(x)$}
\label{expansion}

The integrals $\, I_n(x)$ write as linear combination 
of all the formal solutions at $\,  x=\,  0$ of $\,  {\cal L}^{scal}_q$
\begin{eqnarray}
\label{Inx}
\hspace{-0.95in} && \quad \quad \quad \quad  \quad \quad  \quad  \quad 
I_n(x)\,  \, =\, \,\,  \,  \, 
  \sum_{j=1}^{q} \, c_j^{(n)} \cdot \, S_j^{(n)}, 
\end{eqnarray}
Note that 
the numerical values $\,  c_j^{(n)}$ do depend on the basis chosen 
for the formal solutions $S_j^{(n)}$ (see \ref{I2again} which gives 
the constants for $\,  I_2(x)$ with another combination of formal solutions).
However, as an expansion, $\,  I_n(x)$ is obviously {\em not }
dependent on the basis. For instance, if we trust 
the guessed constants $\,  c_1^{(3)}$ and $\,  c_2^{(3)}$, 
the integral $\,  I_3(x)$ reads: 
\begin{eqnarray}
\label{I3x}
\hspace{-0.95in} && \quad      \quad \,\, 
I_3(x) \,\,=\,\,\,  -{\frac{1}{6 \pi^3}}\, \cdot  \,  
\Bigl(1\,\,  +{\frac{7}{16}} \cdot \, x^2\,
+{\frac{7}{1024}}\cdot \,  x^4\,  \, + \cdots \, \Bigr) \cdot \,  \ln(x)^3
 \nonumber \\
\hspace{-0.95in} && \quad  \quad  \,\,
 \quad  \quad +\, {\frac{1}{6 \pi^3}}
\cdot  \, \Bigl( 3 \, \cdot \, (2\,\ln(2)\,   -\gamma  \, -2)
 \, \, \, + {\frac {21}{16}} \cdot \, (1\,  +2\,\ln(2)\,   -\gamma) \cdot \,   x^2
 \nonumber \\
\hspace{-0.95in} && \quad \quad  \,\,
 \quad  \quad \quad\quad  \quad \quad + {\frac {3}{2048}} \cdot \, (15\,  +28\,\ln(2)\,  
\,   -14\,\gamma) \cdot \,   x^4 \, \,   + \,  \cdots \, \Bigr)\cdot \, \ln(x)^2
 \nonumber \\
\hspace{-0.95in} && \quad  \quad  \quad   \quad  \,\,
-{\frac{1}{6 \pi^3}} \cdot \, 
\Bigl(8.124485 \, + 6.974855 \,{x}^{2}\,  +0.117211 \,{x}^{4}\,  +\cdots \, \Bigr)
 \cdot \,\ln(x)
 \nonumber \\
\hspace{-0.95in} && \quad \quad  \quad  \quad  \,\,
 + {\frac{1}{6 \pi^3}} \cdot \, 
\Bigl(- 7.387058\,  + 7.260657\,{x}^{2}\, 
 +0.150333\,{x}^{4}\, \,   +\,  \cdots\,  \Bigr).
\end{eqnarray}
In front of $\,  \ln(x)^3$, there is the overall constant $\,  c_1^{(3)}$. For 
$\,  \ln(x)^2$, there is no overall constant, because  
the series in front of $\,  \ln(x)^2$ is a sum of two series with the 
combination coefficients $\,  c_1^{(3)}$ and $\,  c_2^{(3)}$.
The same occurs for the others series in front of $\,  \ln(x)$ and $\,  \ln(x)^0$.

For $\,  I_4(x)$, the expansion reads:
\begin{eqnarray}
\label{I4}
\hspace{-0.95in} &&   
\, \,    I_4(x) \, \, \, =\, \, \,   \, \, 
  {\frac{1}{24 \pi^4}} \cdot  \, 
 \, \Bigl(1\, \,  -{5 \over 4}\,{x}^{2}\, \, -{\frac {5}{256}}\,{x}^{4}\,
 \,  -{\frac {1}{2304}}\,{x}^{6} \, \,  + \,  \cdots\, \Bigr) \cdot \, \ln(x)^4
\nonumber \\
\hspace{-0.95in} && \quad \,\,\, \, \, 
  +\,  {\frac{1}{24 \pi^4}}\, \cdot \, 
( 8.76369\,  + 9.29539\,{x}^{2}\,  + 0.143287\,{x}^{4}\,  
+ 0.00397265\,{x}^{6} \,   +\,   \cdots \,) \cdot \, \ln(x)^3 
 \nonumber \\
\hspace{-0.95in} && \quad \,\,\, \, \, 
  +\,   {\frac{1}{24 \pi^4}} \cdot  \,  
 (38.2792 \,  - 39.8374\,{x}^{2}\,  - 0.611859\,{x}^{4}\,  
- 0.0176486\,{x}^{6} +\,   \cdots \,) \cdot \, \ln(x)^2 
\nonumber \\
\hspace{-0.95in} && \quad \,\,\,\, \,  
    + {\frac{1}{24 \pi^4}} \cdot \, 
 (89.0014 \, + 91.2468\,{x}^{2} \, +1.30355\,{x}^{4}  \, + 0.0383979\,{x}^{6}
 + \cdots \, ) \cdot \, \ln(x)
 \nonumber \\
\hspace{-0.95in} && \quad \,\,\, \, \, 
   + {\frac{1}{24 \pi^4}} \cdot \,   
(89.7926\,   - 88.8183\,{x}^{2}\,  - 0.942513\,{x}^{4}\, 
 - 0.0307719\,{x}^{6}\, \,  + \cdots \, ).
\end{eqnarray}

\vskip 0.1cm

{\bf Remark 1:} In the numerical evaluation of the constants $\, c_j^{(n)}$ 
by linear systems like (\ref{I1vert}), the issue of the numerical 
accuracy raises. For the left hand side (\ref{I1vert}) it is 
straigthforward to have the series $\, S_j^{(n)}$ to any length. The 
difficulty is in the numerical evaluation of the multiple integrals 
(\ref{Inintegrals}) which controls the number of digits of the constants $\,c_j^{(n)}$.

\vskip 0.1cm

{\bf Remark 2:} In the evaluation of the linear systems like (\ref{I1vert}), the matching 
point $\, x =  \, x_0$ is used. The value of $\,x_0$ can be any positive number, since the integrals
$\, I_n(x)$ are defined for the positive $\,r=\,x/2$, and since the solutions $\, S_j^{(n)}$ are
given by linear ODEs which have only $\, x=\, 0$ and $\, x=\, \infty$ as singularities.

\vskip 0.2cm

We now turn to the diagonal correlation functions $\,  C (N, \, N)$, 
which write as expansion on the
form factors $\,  f^{(n)}_N$. The linear differential equations 
that annihilate the $\,  C (N, \, N)$ 
are of order $\,  N+1$. \ref{CNNodegeneric} shows that we can find 
the coefficients for generic $\,  N$, but to go further,
a recursion on these coefficients should be found. This 
seems hard to achieve.
Fortunatly, there is a way to produce the linear differential equation
 at scaling that should contain the scaling limit of $\,  C (N, \, N)$.

\section{Painlev\'e VI sigma form equation in the scaling limit}
\label{Painleve}

It is known that the diagonal correlation functions of the Ising model,
$\,C_N =\,\,  C(N,\, N)$ verify the Painlev\'e VI equation 
{\em in its sigma form}~\cite{jim-miw-80}
\begin{eqnarray}
\label{sigma}
\hspace{-0.95in} && \quad \, \,  
 \quad \Bigl(t \cdot \,  (t-1) \cdot \,   {\frac{d^2 \sigma}{dt^2}} \Bigr)^2\,   
\, + 4 \cdot \, 
\Bigl((t-1) \,  {\frac{d \sigma}{dt}}\,  \,  -\sigma -{1 \over 4} \Bigr)
\cdot \, \Bigl( t \cdot \,  {\frac{d \sigma}{dt}}\, \, -\sigma  \Bigr)
 \cdot \, {\frac{d \sigma}{dt}}
\nonumber \\ 
\hspace{-0.95in} && \quad  \quad \quad  \quad \quad  \quad  \quad  \, \,  
 \,\,   =\,  \, \,  \,  
N^2 \cdot \,  \Bigl((t-1) \cdot \, {\frac{d \sigma}{dt}}\,\,  -\sigma \Bigr)^2, 
\end{eqnarray}
where:
\begin{eqnarray}
\label{sigma12}
\hspace{-0.95in} && \quad  \qquad \quad 
\sigma \, \, = \,\, \,   t \cdot \,  (t-1) \cdot  \,   {\frac{d}{dt}}\,   \ln(C_N)\, 
 \, \,  -{t \over 4}, \qquad \qquad  T \,  < \, T_c  \\
\hspace{-0.95in} && \quad \qquad \quad 
\sigma \,\,  = \,\,  \,  t\cdot  \,  (t-1)\cdot  \,   {\frac{d}{dt}}\,   \ln(C_N)\, 
 \,\,   -{1 \over 4}, \qquad \qquad T \,  > \,  T_c 
\end{eqnarray}
The scaling limit of this equation has been given by Jimbo and Miwa~\cite{jim-miw-80}.
It is obtained by simply performing the approriate change of variable, which amounts 
to changing to the variable $\,  x =\,   (1-t) \cdot \, N$ in {\em Painlev\'e VI 
sigma form}, keeping the leading $\,  N$ term.
This gives the scaling equation (irrespective of the regime 
$\,  T < \,  T_c\, $ or $\,\,  T > \,  T_c$)
\begin{eqnarray}
\label{sigmascale}
\hspace{-0.95in} && \quad \quad 
\quad {x}^{2} \cdot \, ({\frac {d^{2} \mu}{d{x}^{2}}})^{2}
\, \, \,   + \,  \, 
\Bigl(4 x  \cdot \, ({\frac {d \mu}{dx}})\, \,  -{x}^{2}\, \,  -4 \mu\Bigr) 
\cdot \, ({\frac {d \mu}{dx}})^{2}
 \, \,  \,  \,  
-{1 \over 2}\,   x \cdot \,  ({\frac {d \mu}{dx}})  \, \cdot \, (1 \, -4 \mu) 
\nonumber \\
\hspace{-0.95in} && \quad \qquad \qquad \qquad \quad 
\,  \,  =  \, \, \,   \,   {1 \over 16}  \, (1 \,  -4 \mu)^{2}, 
\end{eqnarray}
with:
\begin{eqnarray}
\label{mu}
\hspace{-0.95in} && \quad \quad \quad \qquad  \qquad 
\mu \, \,=\, \, \,  \, x\, \,   {\frac{d}{dx}} \,  \ln(C_{scal}(x)).
\end{eqnarray}

To make the expressions closer to the sigma form, one may define 
\begin{eqnarray}
\label{nu}
\hspace{-0.95in} && \quad \qquad  \qquad  \qquad 
\nu   \,  \,=\, \,  \,   \,  
x \cdot \,  {\frac{d}{dx}} \ln(C_{scal}(x))\, \, \,   -{1 \over 4}. 
\end{eqnarray}
Equation (\ref{sigmascale}) becomes
\begin{eqnarray}
\label{sigmascale2}
\hspace{-0.95in} && \quad   \quad 
 {x}^{2} \cdot \,  ({\frac {d^{2} \nu}{d{x}^{2}}})^{2}
\,  \, \, + 4 \, \cdot \, 
  \Bigl(x  \cdot \, ({\frac {d \nu}{dx}})\,\,  -\nu\, \,  -{1 \over 4}\Bigr)
 \, \cdot \,  ({\frac {d \nu}{dx}})^{2}
 \, \,\, \,\, \,= \,\, \, \,\, \, 
(x  \cdot \, ({\frac {d \nu}{dx}})\, \, -\nu)^{2}. 
\end{eqnarray}

\vskip .1cm 

{\bf Remark 3:} 
The form (\ref{sigmascale2}) is equation (38) in~\cite{jim-miw-80}, and 
identifies with (\ref{P5barry}) on $\,  \zeta$ with $\,  r =\,   x/2$.
It seems that this identification between
(\ref{P5barry}) and (\ref{sigmascale2}) (and thus eq.(38) in~\cite{jim-miw-80})
has not been remarked. 

\vskip 0.1cm

{\bf Remark 4:} 
Recall that (\ref{sigmascale2}) is for $\,  C(N,\,  N)$ while (\ref{P5barry}) is for $\,  C(M,\,  N)$. 
The factor $\,  {\cal M}_{\pm}^2$ in (\ref{firstFpm}) is taken care of by the $-1/4$ appearing
in (\ref{nu}). Equation (\ref{sigmascale2}), which is the scaling limit of the equation that
annihilates $\,C(N, N)$, (i.e. (\ref{sigma})) could also be the scaling limit of a 
{\em non-linear} equation (of the Painlev\'e type) for $\, C(M, N)$, 
generalizing (\ref{sigma}), if it exists.

\subsection{Some solutions of the Painlev\'e VI sigma form equation in the scaling limit}
\label{somePainl}

In order to find some of the (non logarithmic) solution $\,  C_{scal}(x)$, 
we plug in (\ref{sigmascale2})
the form  
\begin{eqnarray}
\label{laurentseries}
\hspace{-0.95in} && \quad \qquad \qquad  \,  \, 
C_{scal}(x)  \,\,=\, \, \, \,\,  x^\alpha \cdot \,  \sum a_k \cdot \, x^k, 
\end{eqnarray}
and solves, term by term, on the coefficients $\,  a_k$.  
For generic $\,  \alpha$, one obtains
\begin{eqnarray}
\label{firstCscal}
\hspace{-0.95in} && \quad \qquad \qquad  \,  \, 
C_{scal}^{(I)}(x)  \, \,=\,  \,  \, \,\,
 x^\alpha \cdot \,  
\exp\Bigl(\pm i\, {\frac{4\alpha-1}{8\, \sqrt{\alpha}}} \cdot \,x\Bigr).
\end{eqnarray}

The value $ \, \alpha= \, 1/4$ pops out as particular. When 
fixed and plugging (\ref{laurentseries}) in
(\ref{sigmascale2}), one obtains a one-parameter solution that reads:
\begin{eqnarray}
\label{genCscal}
\hspace{-0.95in} && \quad \quad 
 \quad C_{scal}(x)\,  \,  \,=\,\, \, \, \, \,   x^{1/4} \cdot  \,  
\Bigl(a_0  \, \, \,    + a_2 \cdot \,  x^2\, \, \,  
 + {a_2 \over 64} \cdot  \,  x^4 \,\, \,  
 + \, {\frac{a_2\cdot  \, (a_0 \, -8a_2)}{4608 \,  a_0}} \cdot  \,   x^6
 \nonumber \\
\hspace{-0.95in} && \quad \quad \quad \qquad  \,  \,\, 
\,\, \,  + \, {\frac{a_2 \cdot \,  (5a_0 \, -64a_2)}{2359296 \,  a_0}} \cdot \,   x^8
\, \, \,  +\,  {\frac{a_2  \cdot \,  (7a_0\, -104a_2)}{471859200 \, a_0}} \cdot \, x^{10}
\nonumber \\
\hspace{-0.95in} && \quad  \qquad \qquad  \,  \, 
\,  + \, {\frac{a_2 \cdot \, (21a_0^2\, -296a_0 a_2\, -512a_2^2)}{
271790899200  \, a_0^2}}\cdot \,   x^{12}
\, \, \,   +\,   \cdots
 \Bigr).
\end{eqnarray}

For the value $\,  a_0 =\,  0$, the solution corresponds 
to (\ref{firstCscal}) for  $\,  \alpha =\,  1/4\,  +2$.
For $\,  a_2=\,  0$, the solution is
\begin{eqnarray}
\label{2Cscal}
\hspace{-0.95in} && \quad  \quad  \quad  \quad  \quad \quad  \quad  \quad 
C_{scal}^{(II)}(x)\,\,  =\,\,  \,\, \,  x^{1/4}.
\end{eqnarray}

Now we want to find whether there are particular values of $\,  a_2/a_0$ for which the
series $\,  C_{scal}(x)/x^{1/4}$ in (\ref{genCscal}) verifies a linear ODE. For 
this we use the methods developped 
in~\cite{2008-experimental-mathematics-chi,2009-chi5,2010-chi6,2010-chi5-exact} 
(see also section 6 in~\cite{2014-Ising-SG}, section 3 in~\cite{2014-LGF7})
and consider the series (\ref{genCscal}) modulo a given
prime $\, p_r$. This way, as far as the coefficient $\,  a_2/a_0$ is rational, 
its value is restricted to the interval $\,[1, p_r]$. We then 
let $\, a_2/a_0$ varies over the whole interval $\,  [1, p_r]$ until 
a linear ODE is found. We have written the linear ODE in the (homogeneous) derivative 
$\,  x\cdot \, D_x$, the coefficient-polynomials being of degree $\,  D$,
 and searched for an ODE of order $\, Q \, \le\, 6$, with $\,  (Q+1)\,(D+1) \le \, 220$.
In this range of $Q$ and $D$, there are only the values $\,  a_2/a_0\,  =\,  1/16$ 
and $\,  a_2/a_0=\,  1/32$ that are found, for which the series (\ref{genCscal}) 
is annihilated by a linear ODE.

For the particular value $\,a_2 =\, a_0/16$, the linear 
ODE found is of {\em order two}, with the non logarithmic
solution 
\begin{eqnarray}
\label{3Cscal}
\hspace{-0.95in} && \quad \quad \quad  \quad \quad \quad  \quad  
C_{scal}^{(III)}(x) \,\, \,=\, \, \, \, \, \,
  x^{1/4} \cdot \,  B_0({{x} \over {2}}).
\end{eqnarray}
For the particular value $\,  a_2 =\,  a_0/32$, the linear ODE is of {\em order one}
 and solves as:
\begin{eqnarray}
\label{4Cscal}
\hspace{-0.95in} && \quad \quad \quad \quad \quad   \quad \quad
C_{scal}^{(IV)}(x) \, \,=\, \, \,  \,  \,  \,
 x^{1/4} \cdot \,  \exp\left({\frac{x^2}{32}}\right).
\end{eqnarray}

\subsection{Correspondence with solutions of PVI sigma form equation}
\label{corresp}

In the scaling limit, we have obtained that
Painlev\'e VI sigma form equation has the following solutions:
\begin{eqnarray}
\label{Csca}
\hspace{-0.95in} && \quad  \qquad 
C_{scal}^{(I)}(x)  \,\, = \, \,  \,  x^\alpha \cdot \,  
\exp(\pm i\, {\frac{4\alpha-1}{8\, \sqrt{\alpha}}}\,x), 
\qquad \quad \quad  \alpha \ne\,   1/4, 
\nonumber \\
\hspace{-0.95in} && \quad  \qquad 
C_{scal}^{(II)}(x)  \,\, = \, \,  \, x^{1/4},  
\quad \quad \qquad \quad \quad 
C_{scal}^{(III)}(x)  \, \,= \,\,  \,  x^{1/4}  \cdot \,  B_0({{x} \over {2}}), 
\nonumber \\
\hspace{-0.95in} && \quad  \qquad 
C_{scal}^{(IV)}(x)  \, \,= \,\,  \,  x^{1/4} \cdot \,  \exp\left({\frac{x^2}{32}}\right).
\end{eqnarray}

For the logarithmic solution of the Painlev\'e VI sigma form in the scaling limit, 
the first terms are given in~\cite{mc-tr-wu-77}. More 
terms are given in \ref{nonanal}.

Now, we show the solutions of the Painlev\'e VI sigma form corresponding to these
scaling solutions, $C_{scal}^{(I)}(x), \cdots, C_{scal}^{(IV)}(x)$.
There is one solution to the Painlev\'e VI sigma form which reads
\begin{eqnarray}
C(t)  \, \, \,  \,= \, \,  \, \,\, \,   t^{\alpha} \cdot \, (t-1)^\beta, 
\end{eqnarray}
with (for $\,  T > \,  T_c$)
\begin{eqnarray}
\label{alpha}
\hspace{-0.95in} && \quad  \qquad \qquad 
\alpha  \,\,   = \, \,  \,  \,
 -{1 \over 8}\,  \,   -{1 \over 2}\, \beta \,\,\,\,\, \pm \,\, 
{\frac{4\beta-1}{8 \,\beta}} \cdot \, \sqrt{\beta \cdot \, (\beta-N^2)} , 
\end{eqnarray}
and (for $\,  T < \,  T_c$)
\begin{eqnarray}
\label{alphabeta}
\hspace{-0.95in} && \quad  \qquad \qquad 
\alpha  \,\, =\, \, \,\,  \Bigl({1 \over 2}\, \,   -{1 \over 8 \beta }\Bigr) \cdot  \,
\, (-\beta \,\,\, \pm  \,\sqrt{\beta \cdot \, (\beta-N^2)}). 
\end{eqnarray}

In the scaling limit, the corresponding linear differential operator is (for both
regimes)
\begin{eqnarray}
\label{long}
\hspace{-0.95in} && \quad  \quad \, 
 64\,\beta \cdot \,x^2 \cdot \, D_x^2\,\,\,  \,   -128\,\beta^2 \cdot \,x \cdot \,D_x
\,  \,  \, +(4\beta \, -1)^2 \cdot \,x^2 \, \, +64 \, \beta^2  \cdot \,  (\beta \, +1), 
 \end{eqnarray}
 with solutions
\begin{eqnarray}
\label{c1beta}
\hspace{-0.95in} && \quad \quad \qquad 
 c_1 \cdot \, x^{\beta} \cdot \,
 \exp\Bigl(i\, {\frac{4\beta-1}{8\, \sqrt{\beta}}}\,x\Bigr)\, \, \,\,   
+\,   c_2 \cdot \, x^{\beta} \cdot \,
 \exp\Bigl(-i\, {\frac{4\beta-1}{8\, \sqrt{\beta}}}\,x\Bigr), 
\end{eqnarray}
which are the solutions (\ref{firstCscal}).

The same solution to the Painlev\'e VI sigma form can be seen as given with
$\, \alpha$ being a free parameter, i.e.
\begin{eqnarray}
\label{Ct}
\hspace{-0.95in} && \quad \qquad  \qquad  \qquad 
C(t)  \, \,  \,= \, \, \,\,   t^{\alpha} \cdot \, (t-1)^\beta, 
\end{eqnarray}
with (for $ \,  T >  \,  T_c$)
\begin{eqnarray}
\label{betaone}
\hspace{-0.95in} && \quad  
\beta  \,\,\,\,  = \,\,  \, \,  {\frac{1}{4 \,N^2 \,+16\alpha \,+4}} \cdot \,
 \Bigl( N^2  \,-8\,\alpha^2 \, -2 \,\alpha \,\,\,
  \pm  \,\,(4 \,\alpha \, +1) \cdot \, \sqrt{4 \,\alpha^2 \, -N^2}  \Bigr), 
\end{eqnarray}
and (for $ \,T <  \,T_c$)
\begin{eqnarray}
\label{betatwo}
\hspace{-0.95in} && \quad   
\beta  \,\,  \,=\, \,  \, 
 {\frac{1}{4 \cdot \,(N^2 \, +4 \,\alpha)}} 
\cdot \, \Bigl(N^2 \, -8 \alpha^2 \, +2 \alpha \, \,
 \pm 2  \, \alpha\, \sqrt{(4 \,\alpha^2 \, -1)^2\,  -4 N^2} \Bigr).
\end{eqnarray}

In the scaling limit, the corresponding linear differential operator 
is (for both regimes)
\begin{eqnarray}
\label{Dx2bis}
\hspace{-0.95in} && \quad  \qquad \qquad \qquad \quad \quad 
 16\,\, x^2 \cdot \,D_x^2 \, \, \,\,  +8\,x \cdot \, D_x \, \, \, \, +1, 
\end{eqnarray}
 with solutions
\begin{eqnarray}
\label{c1x1/4}
\hspace{-0.95in} && \quad \qquad \qquad \quad \quad \qquad 
 c_1 \cdot \, x^{1/4}  \,  \, \, \, + c_2 \cdot \, x^{1/4} \cdot \, \ln(x), 
\end{eqnarray}
giving the solution (\ref{2Cscal}).

We have shown in~\cite{2006-painleve6-Cnn} that any combination of
the two solutions of (with $D_t$ the derivative $d/dt$)
\begin{eqnarray}
\label{Lh}
\hspace{-0.95in} && \quad \qquad 
 L_h \,\,  \,=\, \, \,  \,\,    D_t^2 \, \, \, 
 +\Bigl( {1 \over t} \, \,  + {1 \over 2 \, (t-1)} \Bigr) \cdot \, D_t\, 
 \,   \, \,   -{1 \over 4} {N^2 \over t^2}\, 
  \, \,  + {1 \over 16  \,  (t-1)^2}, 
\end{eqnarray}
actually satisfies the Painlev\'e VI sigma form (\ref{sigma}). 
In the scaling limit, the two solutions are
\begin{eqnarray}
\label{Lhsol}
\hspace{-0.95in} && \quad \qquad \quad \quad  \quad  \quad 
c_1 \cdot \,  x^{1/4} \cdot \, B_0\Bigl({{x} \over {2}}\Bigr) 
 \,  \,  \,  \,  
+c_2 \cdot \, x^{1/4} \cdot \, K_0\Bigl({{x} \over {2}}\Bigr), 
\end{eqnarray}
i.e. the scaling solution (\ref{3Cscal}). 
Note that $ \,  x^{1/4}\cdot \,K_0(x/2)\, $ is {\em also a solution of}
 (\ref{sigmascale2}), and $ \,  L_h$ annihilates
 $ \, \,  (1-t)^{1/4} \cdot \, f^{(1)}_N$. 

\section{Scaling limit of the diagonal correlation 
functions $ \,  C(N, \, N)$}
\label{Scalinglimit}

Now, let us show that the scaling solution (\ref{4Cscal})
\begin{eqnarray}
\label{CsalIV}
\hspace{-0.95in} && \quad \quad \quad \qquad \qquad 
C_{scal}^{(IV)}(x)  \,\, = \, \,  \,\, 
 x^{1/4} \cdot \,  \exp\left({\frac{x^2}{32}}\right), 
\end{eqnarray}
corresponds (up to $x^{1/4}$) to {\em an infinite sum} of the scaling limit 
of the $ \,  f^{(j)}_N$,
i.e. this is the scaling solution (analytical at 
$ \,  x= \,  0$) of $ \,  C(N, \, N)$.

We will consider $ \,  f^{(1)}_N$, $ \,  f^{(3)}_N$ and $ \,  f^{(5)}_N$, 
which are solutions
of respectively $ \,  L_2$, $ \,L_4 \cdot L_2$, and $ \, L_6 \cdot L_4 \cdot L_2$.
These linear differential operators are given in~\cite{2007-holonomy}, 
and we call $ \,  L_2^{scal}$, $(L_4^{scal} \cdot L_2^{scal})$ and
$ \,  (L_6^{scal} \cdot L_4^{scal} \cdot L_2^{scal})$ the corresponding scaling
operators.

The function $ \,  C_{scal}^{(IV)}(x)$ expands as:
\begin{eqnarray}
\label{CsalIVx}
\hspace{-0.95in} &&  
{\frac{ C_{scal}^{(IV)}(x) }{x^{1/4}}} \, \, = \, \, \, \,\,
    1 \,  \, \, \, +{x^2 \over 32} \, \, 
+{x^4 \over 2048} \,  \, +{x^6 \over 2^{16} 3} \, \, 
+{x^8 \over 2^{23} 3} \, \,   +{x^{10} \over 2^{28} 15} \, \, 
+ {x^{12} \over 2^{34} 45} \, \,\, \, \,  + \, \,  \cdots  
\end{eqnarray}
The identification will be done on the formal solutions of the scaling
linear differential operators.

With the first terms of the solution of $ \,  L_2^{scal}$
\begin{eqnarray}
\label{S2} 
\hspace{-0.95in} && \quad \quad \quad \quad \quad \quad \quad 
S_2  \, \, \,  =\, \,  \, \,  \,  
  1 \,\,  \, +{x^2 \over 16}\,\, \, 
+{x^4 \over 1024} \, \, +{x^6 \over 147456}   
\, \, \,\, + \,   \cdots,   
\end{eqnarray}
there is only the constant term which matches.

The analytical solution, at $ \,  x = \,  0$, of
 $ \,  L_4^{scal} \cdot L_2^{scal}$ reads:
\begin{eqnarray}
\label{S42a2}
\hspace{-0.95in} && \quad \quad\quad \quad \, 
S_{42} \, \,  =\,  \,  \, \,  \, \, 
  1 \,  \, \,   \, + a_2  \cdot \, x^2 \, \, 
 + {a_2 \over 64} \cdot \,  x^4  \,\,  
 + {1 \over 2^{14}  \cdot \, 3^2} \cdot \,  x^6 \, 
\nonumber \\ 
\hspace{-0.95in} && \quad \quad  \quad \quad  \quad \quad\quad \quad \, 
+\, \Bigl({1 \over 2^{20} \cdot \, 3^2}\, 
 -{a_2 \over 2^{18} \cdot \, 3}\Bigr) \cdot \, x^8 \,\, \,  \, + \,  \cdots
\end{eqnarray}
With the well suited combination ($a_2 =\, 1/32$),  $\, S_{42}$ becomes:
\begin{eqnarray}
\label{S422048}
\hspace{-0.95in} && \quad \quad \quad \quad 
S_{42}  \,   \,=\, \,  \,  \, \,
  1 \,\,\,  \, +{x^2 \over 32} \,\,  \, \,+{x^4 \over 2048}
\,\, \, +{x^6 \over 2^{14} 3^2}\, \,\,  \, + \, \,  \cdots  
\end{eqnarray}
We see that, up to $ \,  x^4$, the coefficients are 
reproduced, i.e. up to $ \,  x^4$, the solution $ \,  C_{scal}^{(IV)}(x)$ 
is reproduced by the scaling limit of:
\begin{eqnarray}
\label{fN1fN3}
\hspace{-0.95in} && \quad \quad \quad \quad \quad \quad \quad
(1-t)^{1/4} \cdot \, \Bigl(f^{(1)}_N  \, \,  + f^{(3)}_N\Bigr).
\end{eqnarray}
Note that the next coefficients of $ \,  C_{scal}^{(IV)}(x)/x^{1/4}$, 
and $ \,  S_{42}$ (i.e. at $ \,  x^6$),
are in the ratio $ \,  4/3$.

Next, we consider the scaling of
\begin{eqnarray}
\label{fN1fN3fN5}
\hspace{-0.95in} && \quad \quad \quad \quad \quad \quad
(1-t)^{1/4} \cdot \, \Bigl(f^{(1)}_N  \,+ f^{(3)}_N \, + f^{(5)}_N\Bigr).
\end{eqnarray}
This amounts to considering the solution of 
$ \,  L_6^{scal} \cdot L_4^{scal} \cdot L_2^{scal}$
\begin{eqnarray}
\label{S642}
\hspace{-0.95in} && \quad
\quad S_{642}\, \,   \,=\, \,  \,\,   \, 
  1 \, \,\, \,  + a_2  \cdot \,  x^2 \,\, \,  + {a_2 \over 64}\cdot  \,   x^4  
\, \, \, +a_6 \cdot \, x^6 \,\,\,   + \, \Bigl({a_6 \over 64}\,
  -{a_2 \over 2^{18} \cdot 3}\Bigr) \cdot \,  x^8 
 \nonumber \\
\hspace{-0.95in} && \quad \quad \qquad
+ \,  \Bigl({13 a_6 \over 2^{15}\cdot 25}\, 
 -{a_2 \over 2^{20}\cdot 75}\Bigr)  \cdot \, x^{10} \,  
 \\ 
\hspace{-0.95in} && \quad \quad
 \qquad + \, \Bigl({49 a_6 \over 2^{18}\cdot 3^2 \cdot 5^2} \,  
 -{11 a_2 \over 2^{26} \cdot 3^4 \cdot 5^2} \,  
-{1 \over 2^{30} \cdot 3^4 \cdot 5^2}\Bigr) \cdot \, x^{12}
 \, \, \,  \,  +\, \cdots, 
 \nonumber 
\end{eqnarray}
and obtaining the well suited combination
 ($a_2= \,  1/32$, $1/a_6= \,  2^{16} \cdot 3$)
\begin{eqnarray}
\label{S642bis}
\hspace{-0.95in} && \, 
 S_{642}   \,=\, \,  
 1 \,  \,\,    +{x^2 \over 32} \, \,   +{x^4 \over 2048}
 \, \,   +{x^6 \over 2^{16} \cdot  3} \,  +{x^8 \over 2^{23} \cdot 3}
 \,\,    +{x^{10} \over 2^{28} \cdot 15} \, \,   
+ {43\,  x^{12} \over 2^{34} \cdot 3^4 \cdot 5^2} \, \,   + \, \cdots  
\end{eqnarray}
which reproduces $ \,  C_{scal}^{(IV)}(x)/x^{1/4}$ up to $ \,  x^{10}$, 
the ratio of the next coefficients
$\, 43/45$ being almost the unity.

With the first three form factors, we may infer that, for
each $ \,  f^{(2n+1)}_N$ form factor added to $ \,  C(N, \, N)$, 
the coefficients of the scaling function 
are reproduced up to $ \,  x^{n(n+3)}$.

Indeed, and as a last check, we consider the next form factor $ \,  f^{(7)}_N$
whose scaling limit is given by
$ \, L_8^{scal} \cdot L_6^{scal} \cdot L_4^{scal} \cdot L_2^{scal}$,
and its analytical solution (at $ \,x= \,0$) which reads:
\begin{eqnarray}
\label{S8642}
\hspace{-0.95in} && \quad
 \quad S_{8642}\,\,  \,= \, \, \, \, \,
  1\,\, \, \, + a_2 \cdot  \, x^2\, \, + {a_2 \over 64} x^4 \,  \,
+ \, a_6 \cdot  \,x^6 \, \, 
 + \, \Bigl({a_6 \over 64}  \, 
 -{a_2 \over 2^{18} \cdot 3}\Bigr) \cdot \, x^8 \,   
 \nonumber \\
\hspace{-0.95in} && \quad \qquad \qquad 
+ \,  \Bigl({13 \, a_6 \over 2^{15} \cdot  25} \, 
-{a_2 \over 2^{20} \cdot  75}\Bigr) \cdot \,   x^{10} 
\, \,  \, + \,  a_{12} \cdot \,  x^{12}  
 \nonumber \\
\hspace{-0.95in} && \quad\qquad 
\qquad   +\, \Bigl({a_{12} \over 64} \,  
 -{33  \, a_6 \over 2^{24} \cdot  7^2 \cdot  5}  \, 
 +{a_2 \over 2^{29} \cdot  3^2 \cdot  5  \cdot 7^2}\Bigr) \cdot \,  x^{14}
 \, \, \,  \, + \,  \cdots
\end{eqnarray}
With the well suited combination $\, a_2= \, 1/32$, $\, 1/a_6= \, 2^{16}\cdot 3$,
$ \, 1/a_{12}= \, 2^{34} \cdot 45$, $\, S_{8642}$ reproduces $\, C_{scal}^{(IV)}(x)/x^{1/4}$ up
to $ \, x^{18}$, and the next coefficients are in the ratio $\, 1571/1575$.

Note that we have the same results when we consider the scaling limits
of $ \,f^{(n)}_N$ with $ \,n$ even. For this, let us show the analytical 
solution, at $\, x=\, 0$, for $ \, L_5^{scal} \cdot L_3^{scal} \cdot L_1^{scal}$,
 which reads
\begin{eqnarray}
\label{S531}
\hspace{-0.95in} && \quad 
 \, S_{531}\,  \,=\, \, \, \,  \, \,
  1\,\,\,  \, + a_2 \cdot \,   x^2\ , \,\,  \, + {a_2 \over 64}\cdot \,    x^4 
\, \,\,   +a_6 \cdot \, x^6\,  \,
+ \, ({a_6 \over 64} \,  
 -{a_2 \over 2^{18} \cdot 3}) \cdot \,    x^8  \\
\hspace{-0.95in} && \, \quad  \quad \, \, 
 \,  + \,  \Bigl({13  \,a_6 \over 2^{15}\cdot 25} \,  
 -{a_2 \over 2^{20}\cdot 75}\Bigr) \cdot \,   x^{10} \, 
 \, \,  + \,  \Bigl({37  \,a_6 \over 2^{18}\cdot 3^2 \cdot 5^2} \,  
 -{a_2 \over 2^{23}\cdot 3^4 \cdot 5^2} \Bigr) \cdot \,   x^{12}
 \, \,  \, \,  + \,  \cdots, 
 \nonumber 
\end{eqnarray}
where we remark that this solution identifies with $ \,  S_{642}$,
 up to $ \,  x^{10}$, i.e.
for the {\em same} well suited combination it reproduces 
$ \,  C_{scal}^{(IV)}(x)/x^{1/4}$ up to $ \,  x^{10}$.
In other words the scaling limit of 
$ \,  f^{(1)}_N \,  + f^{(3)}_N \,  +\cdots \,   +f^{(2n+1)}_N$
identifies with the scaling limit of 
$ \,  1 \,  +f^{(2)}_N \,  + f^{(4)}_N \,  + \,  \cdots \,   +f^{(2n)}_N$,
up to $ \,  x^{n(n+3)}$. As far as the analytical solution at $\, x= \, 0$ of 
the scaling function is concerned, the scaling function is 
the {\em same for both regimes}
(high and low temperatures).

The scaling limit of $ \,  C(N,\,  N)$ is therefore:
\begin{eqnarray}
\label{limit}
\hspace{-0.95in} && \quad \quad \quad \quad 
\lim_{t \rightarrow 1, N \rightarrow \infty}\,
(1-t)^{1/4} \cdot \,\sum_{n: odd, even} f_N^{(n)} (t)
 \,\,\,  = \, \,  \,\,
 x^{1/4} \cdot \, \exp \left( {x^2 \over 32}\right). 
\end{eqnarray}

\section{Scaling limit of the next-to-diagonal 
correlation functions $\,  C(N, \, N+1)$}
\label{AppCNNp1}

The non-diagonal correlation functions $\, C(N, M)$ are given in terms of determinants
(see~\cite{montroll-potts-ward-63}). It has been shown in~\cite{yang-perk-87} that the 
next-to-diagonal correlation functions $\, C(N, N+1)$ have the form of a bordered
Toeplitz determinant.
An iteration scheme of the diagonal and the next-to-diagonal correlation functions
is given by Witte~\cite{witte-07}.

Unlike the diagonal correlation functions $\, C(N,\,  N)$ which are annihilated by 
Painlev\'e VI equation, there is no known (non-linear) 
differential equation for $\, C(N,\,  N+1)$
on which the simple scaling limit 
$\, t \, \rightarrow\,  1$, $\, N \, \rightarrow \, \infty$ can be performed.
However, these next-to-diagonal correlation functions can be written as sum of the
form factors~\cite{Orrick}, $\,C^{(n)}(N,\,  N+1)$. 
In \ref{AppCjNNp1}, we show that these next-to-diagonal form factors are annihilated by
linear ODEs that can be obtained for {\em generic} $\, N$. We 
give in \ref{AppCjNNp1} the first
three linear differential operators
 and their corresponding linear differential operators in the scaling limit.

It appears that these linear differential operators, in the scaling limit, 
identify with the operators for the diagonal $\,  f^{(n)}_N$ in the scaling limit. 
Therefore, we will expect the occurrence of the same expression
$\,  x^{1/4} \cdot \, \exp(x^2/32)$ as the scaling limit of $\,  C(N, N+1)$.

Consider the first term $\,  C^{(1)}(N,\,  N+1)$ whose scaling limit 
is given by the direct sum $\,  L_2^{scal} \oplus L_1^{scal}$,
which has the analytic solution at $\,  x=\,  0$:
\begin{eqnarray}
\label{a0a2a2}
\hspace{-0.95in} && \quad \quad  
a_0 \, \,  \,  \,  + a_2 \cdot \,  x^2\, \, \,   
+ {1 \over 64}\,   a_2 \cdot \,   x^4\, \, \,  \,  
+{1 \over 9216}\,   a_2 \cdot \,  x^6\,   \, 
 + {1 \over 2359296}\,   a_2 \cdot \,  x^8 \, \, \, \, \,    +\,   \cdots
\end{eqnarray}
For $\,  a_0=\,  1$ and $\,  a_2=\,  1/32$, 
there is matching with  $\,  \exp( x^2/32)$ up to $\,  x^4$.

The two terms $\, C^{(1)}(N,\,  N+1) \, +C^{(3)}(N,\,  N+1)$ are annihilated 
by the operator $\, {\cal V}_{10}$ which solves 
$\, C^{(2)}(N,\,  N+1)$ as well. But we have shown in \ref{AppCjNNp1} 
that in the scaling limit, the operator $\, {\cal V}_{10}$ has 
the {\em direct sum} decomposition (\ref{scalV4321}). This allows 
us to pick only the operators 
$\, L_1^{scal} \oplus L_4^{scal} \cdot L_2^{scal}$ 
corresponding to the scaling limit of 
$\, C^{(1)}(N,\, N+1)+C^{(3)}(N,\, N+1)$. 
The analytic solution at $\, x=\, 0$ expands as
\begin{eqnarray}
\label{a0a2}
\hspace{-0.95in} && \quad \quad  \quad 
 a_0 \, \,  \,  \,  + a_2 \cdot \,   x^2\, \, \,   
+ {1 \over 64}\,   a_2 \cdot \,  x^4\, \, 
 \,  +a_6 \cdot \,x^6 \, \,  \,  
+ \, ({a_6 \over 64} \, -{a_2 \over 786432}) \cdot \,   x^8 
\nonumber \\
\hspace{-0.95in} && \quad  \quad  \quad  \quad \quad  \quad 
 + \, ({13 a_6 \over 102400} \, -{a_2 \over 78643200}) \cdot \,   x^{10}
\, \,  \,   \,  +\,   \cdots
\end{eqnarray}
and for $\,  a_0=\,  1,\,   a_2=\,  1/32,\,   a_6=\,  1/196608$, 
matches with $\,  \exp( x^2/32)$ up to $\,  x^{10}$.

Let us consider the whole solutions of the 
scaling limit of $\,  {\cal V}_{10}$, which means that 
we are matching $\,  \exp( x^2/32)$ to the scaling limit of 
$\,  C^{(1)}(N,\, N+1)+C^{(2)}(N,\, N+1)\,  +C^{(3)}(N,\, N+1)$,
i.e. mixing both regimes. The analytic solution at $\,  x=\,  0$ 
of (\ref{scalV4321}) depends on {\em four free
coefficients}, which when fixed to  
$\,  a_0=\,  1,\,   a_2=\,  1/32,\,   a_6=\,  1/196608,\,   a_{12}=\, 1/773094113280$, 
actually matches $\, \exp(x^2/32)$ up to $\, x^{18}$.

Therefore, we have shown that
\begin{eqnarray}
\label{sm1}
\hspace{-0.95in} && \,  \lim_{s \rightarrow 1, N \rightarrow \infty}
\, s^{-1} \cdot \, (1-s^4)^{1/4} \cdot \, 
\sum_{n: odd, even}\,   C^{(n)}(N,\, N+1) \, \, \, = \,\, \,  \,
   x^{1/4} \cdot \, \exp \left({x^2 \over 32} \right).
\end{eqnarray}

\section{Differential Galois groups of the operators in the scaling limit}
\label{decomp}

The equivalence of two properties, namely 
the {\em homomorphism of the operator with its adjoint}, and either the occurrence of a 
{\em rational solution} for the {\em symmetric (or exterior)} square of that operator, 
or the  drop of order of these squares\footnote[1]{The order of 
the symmetric (or exterior) of these operators is less than the order
generically expected for these squares. In terms of differential systems
this corresponds, however, to rational solutions.},
have been seen for many linear differential operators~\cite{2014-DiffAlg-LGFCY}.

The linear differential operators with these properties are such that their differential Galois 
groups are included in the symplectic, or orthogonal, differential groups. 

The solutions of the operators $\, L_n$, when $\,N$ is fixed to an integer, write
as {\em polynomials} in the complete elliptic integrals $\,K$ and $\,E$. 
The operators are equivalent to some symmetric power of $\,L_E$, the linear 
differential operator for the complete elliptic integral $\,E$.
The homomorphisms of the $\,L_n$ with their corresponding adjoint is, therefore, 
a straightforward consequence of the homomorphism of $\,L_E$ with its adjoint.

Furthermore, we forwarded, in a recent paper~\cite{2014-Decomp-Special},
 a ``canonical decomposition'' for those operators 
whose {\em differential Galois groups}
 are included in {\em symplectic or orthogonal}  groups.
These linear differential operators are homomorph to their adjoints, 
and a ''canonical decomposition'' of these linear differential 
operators can be written 
in terms of a ``tower of intertwiners''~\cite{2014-Decomp-Special}.

The issue, we address in the sequel, is whether these properties hold for
the operators $\, L_n$ with a {\em generic parameter} $\, N$, and whether this 
is preserved in the scaling limit.

We find that the linear differential operators $\, L_n$ (given up to $ \, n= \, 10$ 
in~\cite{2007-holonomy})
are homomorphic to their respective adjoints for {\em generic values} of $\,N$. Their
differential Galois groups are in  {\em symplectic groups} for $\, n$ even,
 and in {\em orthogonal groups} for $\,n$ odd. Their exterior (for $\,n$ even)
 and symmetric (for $\,n$ odd) squares do annihilate a {\em rational 
function}. For instance, the rational solution of the {\em symmetric 
square} of $\,L_3$ reads
\begin{eqnarray}
\label{solRsym2L3}
\hspace{-0.95in} && \qquad \qquad 
sol_R( sym^2(L_3)) \,\, \, = \,\,  \,  \,  \, 
{\frac{N^2 \cdot \,t^2 \, \, 
 - (2\, N^2\, -1)\cdot \,t \,  \, +\, N^2}{t^2 \cdot \,(1-t)^2}}, 
\end{eqnarray}
and the rational solution of the {\em exterior square} of $\,L_4$ is:
\begin{eqnarray}
\label{solRext2L4}
\hspace{-0.95in} && \qquad \qquad 
sol_R( ext^2(L_4)) \,\, \,  \, = \,\, \, \, \,   \, 
  {\frac{(N^2\, -1)\cdot \,t^2\, \,  - 2 N^2 \cdot \,t \,  \, 
+(N^2\, -1)}{t^5 \cdot \, (1-t)^3}}.
\end{eqnarray}

The operator $\, L_3^{scal}$ is the scaling limit of $\, L_3$,
 and is (non-trivially) {\em homomorphic to its adjoint}.
The rational solution of its {\em symmetric square} is
\begin{eqnarray}
\label{solRL3}
\hspace{-0.95in} && \qquad \qquad \quad 
sol_R(sym^2(L_3^{scal})) \,\,= \,\,\,\, \,  {\frac{1\, +x^2}{x^2}}, 
\end{eqnarray}
which is the rational function (\ref{solRsym2L3}) in the scaling limit.

Similarly, the operator $\, L_4^{scal}$ (the scaling limit of $\, L_4$) 
is homomorphic to its adjoint,
and the rational solution of its {\em exterior square} reads
\begin{eqnarray}
\label{solRL4}
\hspace{-0.95in} && \qquad \qquad  \quad 
sol_R(ext^2(L_4^{scal})) \,  \,\,= \,\,\, \,  {\frac{x^2\, -2}{x^3}},
\end{eqnarray}
which, in the scaling limit, is the rational function (\ref{solRext2L4}).

The orthogonal (resp. symplectic) differential Galois groups admit an
invariant quadratic (resp. alternating) form. Here also, for instance for $L_3$,
one has the following quadratic form, depending on $\, N$, 
$\,Q(X_0, X_1, X_2) \,=\, const.$ where
\begin{eqnarray}
\label{quadra}
\hspace{-0.95in} && \,\,
 Q(X_0, X_1, X_2) \,\, = \,\,\,
 \Bigl({t}^{2}   \cdot \, (5 -10\,t+4\,{t}^{2}) \,
 - (1-t)^{4} \cdot \,{N}^{2}  \Bigr) \cdot \,  X_0^{2} \, 
 \nonumber \\
\hspace{-0.95in} && \quad
+ \Bigl({t}^{2} \, (1-t)^{2} 
\, (4-17\,t+16\,{t}^{2})\, 
 -{t}^{2}  \cdot \, (1-t)^{4}\cdot \, {N}^{2}  \Bigr) \cdot \,  X_1^{2}  
\nonumber \\
\hspace{-0.95in} && \quad  
+  \, {t}^{4}  \cdot \, (1-t)^{4} \cdot \, X_2^{4} \,  \,   \, 
- \, \Bigl(2 \,t \cdot \, (1-t)^{4} \, {N}^{2} \, 
+{t}^{2} \cdot \, (1-t)  \, (9-27\,t+16\,{t}^{2})\Bigr)  \cdot \,  X_0\, X_1 
\nonumber \\
\hspace{-0.95in} && \quad
-4 \,{t}^{3}  \cdot \, (1-t)^{3} \cdot \, X_2\, X_0 \, \, 
+4 \,{t}^{3}  \cdot \, (1 -2\,t)  \, (1-t)^{3} \cdot \, X_2\, X_1, 
\end{eqnarray}
which, in the scaling limit, becomes the quadratic form
\begin{eqnarray}
\label{quadrascal}
\hspace{-0.97in} && \quad \quad
\quad Q^{scal}(X_0, X_1, X_2) \,\, \,= \,\,\,\, X_0^{2}\, \, \, 
- {x}^{2} \cdot     \, (3-{x}^{2}) \cdot \,  X_1^{2} \,  \, -{x}^{4}\cdot \,  X_2^{2}
\nonumber \\
\hspace{-0.95in} && \quad \quad \quad \quad \qquad \, \qquad
\, -4\,{x}^{3}\cdot \, X_1\, X_2 \, \,  +2\, x \cdot \,X_0\, X_1, 
\end{eqnarray}
for $ \,L_3^{scal}$. In (\ref{quadra}), and (\ref{quadrascal}), 
$ \,X_0$ denotes any solution of the considered linear differential operator,
$ \,X_1$ and $ \,X_2$ being the first and second derivative of $ \, X_0$.
 
\vskip 0.1cm

The operators $\, L^{scal}_n$ ``inheritate'' the 
differential Galois groups of the operators $\,L_n$.
For $\,n$ even (resp. $\, n$ odd), the differential Galois group of $\,L_n^{scal}$ is 
{\em included} in $\, Sp(n, \, \mathbb{C})$ (resp. $\, SO(n, \, \mathbb{C})$).
Recall that the solutions of the operators $\, L_n$ (resp. $\,L_n^{scal}$) write as
polynomials in the complete elliptic integrals (resp. modified Bessel functions),
which means that the linear differential operators $\,L_n$ (resp. $\,L_n^{scal}$) are 
homomorphic\footnote[2]{$\, L_n$ is homomorphic to the symmetric $\,(n-1)$-th power of $\,L_2$,
with $\, N$ generic (not necessarily an integer).}
to the symmetric $\,(n-1)$-th power of $\, L_2$ (resp. $\,L_2^{scal}$).
Thus, the differential Galois group of $\,L_n$ and $\,L_n^{scal}$ is, in fact, the 
differential Galois group of $\,L_2$ (or $\,L_2^{scal}$), 
namely\footnote[1]{$\, SL(2, \, \mathbb{C})$ is isomorphic to $Sp(2, \, \mathbb{C})$, 
to $Spin(3, \, \mathbb{C})$, and isomorphic, up to a 2-to-1 homomorphism,
to $\, SO(3, \, \mathbb{C}) \simeq \, PSL(2,\, \mathbb{C})$.} 
$\, SL(2, \, \mathbb{C})$.

\vskip 0.1cm

We have shown in~\cite{2014-Decomp-Special} that the homomorphism of the operator 
with its adjoint implies a "canonical decomposition" 
in terms of {\em self-adjoint operators}.
This decomposition is obtained by a sequence of Euclidean 
right divisions (see~\cite{2014-Decomp-Special}
 and section 9 in~\cite{2014-LGF7}).
The operator $\, L_3$ has the canonical decomposition (for generic $\, N$)
\begin{eqnarray}
\label{decomL3}
\hspace{-0.95in} && \qquad \qquad  \quad 
L_3 \,\, \, = \,\,\, \, \,  
(U_1^{(3)} \cdot U_2^{(3)} \cdot U_3^{(3)}\,  + U_1^{(3)} + U_3^{(3)}) \cdot \, r_1^{(3)}(x), 
\end{eqnarray}
where $\, r_1^{(3)}(x)$ is a rational function,
 and where $\, U_1^{(3)}$, $\, U_2^{(3)}$ and $\,U_3^{(3)}$ 
are {\em order-one self-adjoint} operators.
In the scaling limit, one obtains for $\, L_3^{scal}$:
\begin{eqnarray}
\label{decomL3scal}
\hspace{-0.95in} && \qquad \qquad  \quad 
L_3^{scal} \,\, \, = \,\, \, \, \, 
\, W_1^{(3)} \cdot W_2^{(3)} \cdot \, W_3^{(3)} \, 
+ W_1^{(3)} \, + W_3^{(3)}) \cdot \, r_2^{(3)}(x).
\end{eqnarray}
Here also, $\,r_2^{(3)}(x)$ is a rational function,
 and $\, W_1^{(3)}$, $\, W_2^{(3)}$ and $\, W_3^{(3)}$ 
are {\em order-one self-adjoint} operators.

Similarly, the operator $\, L_4$ has the following canonical 
decomposition (for generic $\, N$)
\begin{eqnarray}
\label{decomL4}
\hspace{-0.95in} && \qquad \qquad  \quad 
L_4 \,\, \, = \,\, \, \,  \, 
(U_1^{(4)} \cdot\,  U_2^{(4)}\,  + 1) \cdot \, r_1^{(4)}(x), 
\end{eqnarray}
where $ \, r_1^{(4)}(x)$ is a rational function, and where 
$ \, U_1^{(4)}$ and $ \, U_2^{(4)}$ 
are {\em order-two self-adjoint} operators.
In the scaling limit, one obtains for $ \, L_4^{scal}$
\begin{eqnarray}
\label{decomL4scal}
\hspace{-0.95in} && \qquad \qquad  \quad 
L_4^{scal} \,\,\,  = \,\, \,\,  \,   \,
 (W_1^{(4)} \cdot  \, W_2^{(4)} \,  + 1) \cdot \,  r_2^{(4)}(x),
\end{eqnarray}
where $ \, r_2^{(4)}(x)$ is a rational function, and where 
$ \, W_1^{(4)}$ and $ \, W_2^{(4)}$ 
are {\em order-two self-adjoint} operators.

\vskip 0.1cm

The ``canonical'' decomposition occurring for the operators $\,L_n$, 
is {\em preserved in the scaling limit}. 
In particular the {\em self-adjoint}
operators of these ``canonical'' decompositions~\cite{2014-Decomp-Special} 
are {\em all} of {\em order one} for the  $\,L_n$ and $\,L_n^{scal}$ with $\, n$ odd
and are {\em all} of {\em order two} for the  $\,L_n$ and 
$\, L^{scal}_n$ with $\, n$ even.
The rational solutions of the symmetric, or exterior, 
squares of the $\,L_n^{scal}$ are given in \ref{ratioscal}.

\section{Conclusion}
\label{concl}

To obtain the expression $\, x^{1/4} \cdot \, \exp(x^2/32) \, $ 
as the scaling limit of the correlation functions $\, C(N,\, N)$, we 
have made a ``matching'', in the scaling limit, of both hand-sides of:
\begin{eqnarray}
\hspace{-0.95in} && \qquad \quad  \qquad \quad \, 
C(N, N)  \,\, \, = \,\,\, \, \, 
(1 \, - t)^{1/4} \cdot \, \sum_j \, f^{(j)}_N.
\end{eqnarray}
The left-hand-side is taken as a particular solution that pops out 
from the {\em sigma form} of Painlev\'e VI in the scaling limit. 
The right-hand-side is a particular combination of the sum of the 
(non logarithmic) formal solutions of the operators (annihilating 
$\,f^{(j)}_N$) at scaling.

For the next-to-diagonal correlation functions $\, C(N,\, N+1)$, there 
is no (non-linear) differential equation one can use, but we have obtained that the 
next-to-diagonal form factors $\,C^{(j)} (N, \, N+1)$ have, in the scaling limit,
the {\em same} linear differential operators $\, L^{scal}_n$.  
One may conjecture that we will obtain the same linear differential operators at scaling 
for the $\,j$-contributions $\,C^{(j)}(N,\, N+p)$, 
$\,C^{(j)}(N,\, p \cdot\, N)$ with $\,p >\,1$
or $\,C^{(j)}(N,\, M)$. 

\vskip 0.1cm

 Each time the discrete parameter $\, N$ of the lattice
appears explicitly in a differential equation, the scaling limit
can easily be performed.
The correlation functions $\, C(N, \,N)$ is a solution of the sigma form of
Painlev\'e VI (see (\ref{sigma})) which, itself, is a specialisation of
a more general nonlinear differential
equation~\cite{okamoto-87,forrester-witte-04}, also
called sigma form of Painlev\'e VI, which depends on {\em four}
parameters\footnote[1]{The general Painlev\'e VI sigma form 
(eq.(1) in~\cite{2006-painleve6-Cnn}), deals with
the function $\, \zeta(t)$ and depends on {\em four} parameters $v_1,\,  \cdots, \, v_4$.
Equation (\ref{sigma}) for the $\,C(N,N)$ is the subcase, $\sigma(t) = \zeta(t) +N^2\cdot \, \,t/4-1/8$,
$v_1=\, v_4=\, N/2$, $v_2=\, (1-N)/2$ and $v_3=\, (1+N)/2$.}.
The scaling limit performed on (\ref{sigma}) with (\ref{sigma12}) has given
the nonlinear equation (\ref{sigmascale2}) that identifies with
(\ref{P5barry}) which concerns the scaling limit of the correlation functions $\, C(M, N)$.
If one assumes that, similarly to $\, C(N, \,N)$, the $\, C(M, N)$ also
verify a non-linear differential equation, generalizing (\ref{sigma}), one
possible scenario could be that a two-parameter  nonlinear equation for $\, C(M, N)$
emerges as a subcase of the four-parameter sigma form of
Painlev\'e VI.
Finding this two-parameter  nonlinear equation for $\, C(M, N)$
essentially requires to generalize the definitions of $\, \sigma$, 
namely (\ref{sigma12}), and  to find the constraints on the four parameters.

The square Ising model has shown an extremely rich structure 
illustrated by a large set of exact results corresponding to
highly selected linear differential equations of the 
$\, n$-particle contribution to the magnetic susceptibility $\,\chi^{(n)}$, 
correlation functions $\,C(N,\,  M)$, form factors $\, C^{(j)} (N, \,M)$, etc. 
For the linear ODE 
 which have only the three\footnote[3]{This is at contrast with, for example, the case of the 
magnetic susceptibility of the Ising model which is an {\em infinite sum} 
of contributions with large set of regular singularities that eventually densify the 
unit circle $\, \vert s \vert = \, 1$ yielding a {\em natural boundary}. For 
the scaling function of the magnetic susceptibility $\, \chi$, see~\cite{chan} 
and references therein.} regular singularities $\,t=\, 0$, $\,t=\, 1$
 and $\,t=\, \infty$,  the scaling limit leads to a
{\em confluence}~\cite{Ramis,Ramis2}  of the singularities, ending in the 
regular $\,x= \, 0$ and the {\em irregular} $\,x= \, \infty$ points. 

All the remarkable structures discovered in previous papers,
on the square Ising model (elliptic functions, modular forms, 
Calabi-Yau equations, ``special'' differential Galois groups, 
globally bounded series, diagonals of
rational functions, ...) emerge in a framework
 related to the (Yang-Baxter) integrability concept occurring 
 {\em on a  lattice}. In the scaling limit, with the emergence of 
irregular singularities from the confluence of regular ones, 
 many of these structures actually disappear, 
or are less crystal clear.
For instance, the property of global nilpotence, occurring
 in all our linear ODEs, disappear in the scaling limit,
but some structures still show up for the $\, p$-curvature
(see section 10 in~\cite{2009-global-nilpotence}). In contrast, 
we have seen that the differential Galois group
structures are more robust, being preserved by the scaling limit.

What happens in the scaling limit to
{\em all} the remarkable holomic or non-holonomic structures 
we have discovered in the last decade, on the square Ising model?

\vskip 0.2cm

\ack This work has been performed without any support of the ANR, the ERC or the MAE. 

\vskip .4cm

\appendix

\section{Recall: $\, C(N, \, N)$ and $\, f^{(n)}_N$ as polynomials in $\,K$ and $\,E$}
\label{CnnETfjn}

The correlation functions $\,C(N,\, N)$ are the analytical (at 0) solutions of
linear ODE of order $\,N+1$. For $\,N$ fixed to an integer, the correlation
functions $\, C(N,\, N)$ writes as {\em polynomials} in the complete elliptic integrals
of first and second kind $\,K$ and $\,E$ of homogeneous degree $\,N$.
With
\begin{eqnarray}
\hspace{-0.95in} && \quad \qquad \, 
K \,\, = \, \,\, _2\,F_1([1/2, 1/2], [1], \,t), 
 \quad \, \,\, \, \,\, \, \, 
E \,\, = \, \,\, _2\,F_1([1/2, -1/2], [1],\, t), 
\end{eqnarray}
the form of $\,C(N,\, N)$ reads 
\begin{eqnarray}
\hspace{-0.95in} && \quad \qquad  \quad \quad
C(N,\, N) \,\, = \, \,\,\, \, 
\sum_{i=0}^N \, Q(N, i, t) \cdot \, K^{N-i} \cdot \, E^i, 
\end{eqnarray}
where $\,Q(N,\, i, \,t)$ is a rational function.
For instance $\, C(2,\, 2)$ in the $\,T\, > \,T_c$ regime, writes:
\begin{eqnarray}
\label{C22}
\hspace{-0.95in} && \quad
3 \cdot \,  t \cdot \, C(2,\, 2) \,\,\, = \, \,  \,  \, \,  
3 \cdot \,(t-1)^2 \cdot \, K^2 \,\,    \,
+ 8 \cdot \, (t-1)\cdot \, K \cdot \,E\, \, \,
 - (t-5) \cdot \, E^2.
\end{eqnarray}

The form factors $\, f\, ^{(n)}_N$ are the analytical 
(at $\, 0$) solutions of linear differential operators  with $\, N$ as a parameter.
With $\, n$ and $\, N$ fixed to integers, $\, f^{(n)}_N$ 
writes as a {\em sum of polynomials}
in $\, K$ and $\, E$. 
The form of $\, f^{(2 n+1)}_N$ reads
\begin{eqnarray}
  \label{f2nN}
\hspace{-0.95in} && \quad  \quad \qquad\, 
f^{(2 n+1)}_N \,\, = \, \,\, \,  \, 
 \sum_{j=0}^n \, \sum_{i=0}^{2 j+1} \, 
P(N,\,  n,\,  j,\,  i,\,  t) \cdot \, K^{2 j+1-i}  \cdot \, E^i, 
\end{eqnarray}
with $\, P(N,\,  n,\,  j,\,  i,\,  t)$ a rational function.
In the expression of $\, f^{(2 n+1)}_N$, the homogeneous degrees 
of $\, K$ and $\, E$ occur as $\, 1,\,  3, \, \cdots, \, 2 n+1$. 
Recall~\cite{2007-holonomy} that the linear differential operators 
annihilating the $\, f^{(2 n+1)}_N$, have a 
{\em direct sum} structure
when the parameter $\, N$ is fixed to an integer.
The first two $\, f^{(2 n+1)}_N$ contributing to the example of 
$\, C(2, \, 2)$ are:
\begin{eqnarray}
\hspace{-0.95in} && \,  \quad 
3 \, t \cdot \, f^{(1)}_2  \,\, = \, \, \, \,\, 
 t \cdot \, (t+2) \cdot \, K \, \,\,  - 2\,  t \cdot \, (t+1) \cdot \, E, 
\end{eqnarray}
\begin{eqnarray}
\hspace{-0.95in} && \quad  \, 
18 \, t \cdot \, f^{(3)}_2  \,\, = \, \, \,  \,\, 
-3 \cdot \,  ({t}^{2}-2)\cdot \,   {K}^{3} \, \,  \,
+3 \cdot \, (2\,{t}^{2}-11\,t+2)\cdot \,  {K}^{2} \cdot \,  E
 \nonumber \\
\hspace{-0.95in} && \quad \quad    \quad 
+36\, \cdot \,  (t-1) \cdot \,   K\cdot \,  {E}^{2} \,  \,
+24\,{E}^{3}  \,\,   \, +7 \cdot \, (t+2) \cdot \,   K \, \,
 -14 \cdot \, (t+1) \cdot \,   E.
\end{eqnarray}

The expression (\ref{CNdefplus}) reproduced here for $\,N=\,2$
\begin{eqnarray}
\hspace{-0.95in} && \quad  \qquad  \qquad \quad  \quad  
C(2,\, 2)  \, \,\,  =\,\, \, \,  \,
 (1-t)^{1/4} \cdot \,  \sum_{n=0}^\infty  \, f^{(2n+1)}_2, 
\end{eqnarray}
shows that an {\em infinite sum}
 of polynomials in $ \, K$ and $ \, E$ will give birth
to the overall factor $ \, (1-t)^{-1/4}$ absent in (\ref{C22}).
This situation has been encountered in the magnetic susceptibility of Ising
model at scaling (see section 7 in~\cite{2005-connection-chi3-chi4}).
See also section 5.1 in \cite{2007-diagonal-chi}, where a sum of terms,
each term being a polynomial expression of the complete elliptic integrals,
reduces to an algebraic expression.

\section{Recall of the expressions of  $\,  L_n^{scal}$, $ \, n=\,  1, 2, \cdots, 6$}
\label{Lnscal}

The form factors $\,  f^{(1)}_N$ and $\,  f^{(3)}_N$ are annihilated by the
 order-six operator $\,  L_4 \cdot L_2$, 
which, in the scaling limit, writes $\,  L_4^{scal} \cdot \, L_2^{scal}$, where:
\begin{eqnarray}
\label{L2scal1}
\hspace{-0.95in} &&  \quad 
 \qquad L_2^{scal}  \,\,  = \,\, \, \,  
 4\, x \cdot \,  D_x^{2}\, \,\,  +4 \, D_x \,\,\,  -x, 
\nonumber  \\
\hspace{-0.95in} &&   \quad 
 \qquad L_4^{scal}  \, = \, \,    \,   
 16\,   x^3 \cdot \,   D_x^4\, \, \,  +160\,   x^2 \cdot \,   D_x^3 
\, \, \,  -8 \,  x \cdot \,(5 x^2-46)\cdot \,  D_x^2\, \, 
\nonumber  \\
\hspace{-0.95in} &&  \qquad  \qquad \qquad \quad  \,  \,     
 -72 \cdot \, (x^2\, -2) \cdot \, D_x\, \, \,  +9\,  x^3.
\end{eqnarray}

The form factor $\,  f^{(5)}_N$ is annihilated by 
the order-twelve linear differential operator 
$\,  L_6 \cdot\,  L_4 \cdot \,  L_2$, 
which, in the scaling limit, writes 
$\, L_6^{scal} \cdot \,   L_4^{scal} \cdot\,   L_2^{scal}$, 
where $\, L_6^{scal}$ reads:
\begin{eqnarray}
\label{L6scal}
\hspace{-0.95in} && \,
 \quad L_6^{scal} \, = \,\,\, \,   
64\,{x}^{5} \cdot  \, D_x^{6}  \,\,\, +2240\,{x}^{4}\cdot  \,  D_x^{5} \, \, \, 
-112\,{x}^{3} \cdot \,  (5\,{x}^{2} -236) \cdot \,  D_x^{4}  
\nonumber \\
\hspace{-0.95in} && \,\qquad \quad  \, \,      
 -32\,{x}^{2} \cdot \, (259\,{x}^{2}-3916) \cdot \,   D_x^{3}    \, 
+4 \,x \cdot \, (259\,{x}^{4} \, -7668\,{x}^{2}+54128) \cdot \, D_x^{2}
\nonumber \\ 
\hspace{-0.95in} && \, \qquad \quad  \,   \,    
\,+100 \cdot \, (784 \, -236\,{x}^{2} \, +27\,{x}^{4})  \cdot \, D_x \,
 \, -225\,{x}^{5}.  
\end{eqnarray}

The form factor $\,  f^{(2)}_N$ is annihilated by the order-four operator 
$\,  L_3 \cdot L_1$, which, in the scaling limit, 
writes $\,  L_3^{scal} \cdot L_1^{scal}$, where:
\begin{eqnarray}
\label{L1scal1}
\hspace{-0.95in} && \quad \quad 
L_1^{scal}  \, \, = \,\, \,  \,D_x,
 \nonumber \\   
\hspace{-0.95in} && \quad  \quad 
L_3^{scal}  \,\,  = \,\, \, \,  \, 
2\,{x}^{3} \cdot \,  D_x^{3}\,\,\, \,  \,   +8\,{x}^{2} \cdot \,  D_x^{2}\, \, \,  \, 
-2\,  (x-1)  \, (x+1)\cdot \, x \cdot \, D_x \,\,\,   \,  \,-2.
\end{eqnarray}
Note that $ \, L_3^{scal} \cdot\,  L_1^{scal}$ 
{\em has a direct sum decomposition}
$\, \,   L_3^{scal} \cdot L_1^{scal}\,  =\, \,   L_1^{scal} \oplus \tilde{L}_3^{scal}$,
with
\begin{eqnarray}
\tilde{L}_3^{scal} \,\, \,  = \, \, \, \, \,  \, 
x^2 \cdot \, D_x^3\, \,   \,  + 3 \, x  \cdot \, D_x^2 \,  \, 
+ (1-x^2) \cdot \, D_x \, \,  + x.
\end{eqnarray}

The form factor $\,  f^{(4)}_N$ is annihilated by the order-nine operator 
$\,  L_5 \cdot L_3 \cdot L_1$, 
which, in the scaling limit, writes 
$\, L_5^{scal} \cdot L_3^{scal} \cdot L_1^{scal}$, 
where $\, L_5^{scal}$ reads:
\begin{eqnarray}
\label{L5scal}
\hspace{-0.97in} &&   
 L_5^{scal} \,\, = \,\,\,
2\,{x}^{5} \cdot \, D_x^{5}\,  \,  +40\,{x}^{4} \cdot \,  D_x^{4}\,
  -2\,{x}^{3} \cdot \, (5 \,{x}^{2}-113) \cdot \,  D_x^{3} \, 
-2\,{x}^{2} \cdot \, (32\,{x}^{2}-161)  \cdot \,  D_x^{2}
 \nonumber \\
\hspace{-0.95in} && \quad   \quad \qquad \,
+2\,x \cdot \, (4\,{x}^{4}-97-24\,{x}^{2})\cdot \,  D_x
\,\,\, +32\,{x}^{2}\, -256.   
\end{eqnarray}

\section{$I_2(x)$ again}
\label{I2again}

The choice of the basis of the formal solutions is arbitrary. Instead of the basis 
$\, ( S_1^{(2)}, \, S_2^{(2)},  \, S_3^{(2)}, \,  S_4^{(2)})$, one may take
\begin{eqnarray}
\label{S12S22}
\hspace{-0.95in} && \quad \quad \,
\tilde{S}_1^{(2)} \, \, = \,\,  \, \,  \, 
 S_1^{(2)}\, \,  \,  + S_2^{(2)}\,\,   -{15 \over 2}\,  S_3^{(2)}\,  \, 
 + {17 \over 2}\,  S_4^{(2)}, \\
\hspace{-0.95in} && \quad \quad \,
\tilde{S}_2^{(2)} \, \, = \,\,  \, \,  \, 
 S_2^{(2)}  \,  \, +{39 \over 16}\,  S_3^{(2)} \, \,  - {31 \over 16} \, S_4^{(2)}, 
\quad \quad   \,  \,  \, 
\tilde{S}_3^{(2)} \, \,  \,=  \,\,\,  S_3^{(2)}, 
 \quad \quad \, \,  \, 
\tilde{S}_4^{(2)} \,\,  = \,\, \,  S_4^{(2)}, 
\nonumber 
\end{eqnarray}
where the series begin, now, as $\, const.\, +\cdots$
The combination coefficients $\, \tilde{c}_j^{(2)}$ will appear as 
\begin{eqnarray}
\label{c12c32c}
\hspace{-0.95in} && 
 \quad \tilde{c}_1^{(2)}\,  \,=\,\,\,   0.1013211, 
\quad  \,  \, \tilde{c}_2^{(2)} \,\, =\,\,\,   -0.06263, \quad \,  \, 
\tilde{c}_3^{(2)} \,\, =\,\, \, 0.61863, 
\quad  \,  \, \tilde{c}_4^{(2)} \, \,=\,\, \, -0.53296,   
 \nonumber 
\end{eqnarray}
and in exact forms as:
\begin{eqnarray}
\label{c12c32c4}
\hspace{-0.95in} && \qquad 
\tilde{c}_1^{(2)} \,\, = \, \, \, {1 \over \pi^2}, 
\qquad \qquad \quad 
\tilde{c}_2^{(2)} \,=\, \, 
{1 \over \pi^2}\cdot \, (1\,-4\ln(2)\, +2 \gamma),
 \nonumber \\
\hspace{-0.95in} && \qquad 
\tilde{c}_3^{(2)} \,\, = \,\,\, 
{1 \over 4 \pi^2} \cdot \, (1\,-4 \ln(2)\,+2 \gamma)^2 \,  \,  \, 
+ {1 \over 8 \pi^2}\cdot \, (23+62 \ln(2)-31 \gamma),
 \nonumber \\
\hspace{-0.95in} && \qquad 
\tilde{c}_4^{(2)} \, \,= \,\,\,
 -{1 \over 8 \pi^2} \cdot \,  (17\,+62 \ln(2)\,-31 \gamma).
\end{eqnarray}

\section{ The $\, C(N,\,  N)$ correlation functions}
\label{CNNodegeneric}

The correlation functions $\, C(N,\,  N)$ are annihilated by a linear ODE of order
$\, N+1$. The form of the linear differential operators is
\begin{eqnarray}
\label{Lq}
\hspace{-0.95in} && \quad \quad \quad \quad \quad 
L_{N+1} \,  \,\,=\,\,\, \, \,   
P_{N+1} \cdot \, D_x^{N+1} \,\,   + P_{N} \cdot  \, D_x^{N} \,\, \,  + \, \,  \cdots\,\, \,  + \, P_0, 
\end{eqnarray}
where, for generic $\, N$, the first polynomials $\, P_{N-k}$ (for $\, N >\, k$) read:
\begin{eqnarray}
\label{qN+1}
\hspace{-0.95in} && 
  P_{N+1} \, \, = \,\, \,  \, x^{N+1} \cdot\, (x-1)^N, 
\nonumber \\
\hspace{-0.95in} && 
P_{N} \, = \,\,  -{1 \over 6}  \cdot \, 
x^{N}\, (x-1)^{N-1} \cdot \, \, N \cdot  \, (N+1) \cdot \, \Bigl((N-4) \cdot \, x \, +(N+2) \Bigr), 
\nonumber \\
\hspace{-0.95in} && 
  P_{N-1} \, =\,\, {1 \over 260} \cdot \,   x^{N-1} \cdot \, (x-1)^{N-2} \cdot \, N  \,(N+1)
\Bigl((N-1)(N-2) \cdot \, (5N^2\, -26N\, +18) \cdot \,x^2 
\nonumber \\
\hspace{-0.95in} && 
 \qquad \quad +(N+2) \cdot \, (10N^3 \, -54N^2\, +62N\, -3) \cdot \,x \, 
\nonumber \\
\hspace{-0.95in} &&  \quad \qquad 
+ (N+2)\, (5N^3\, +9N^2\, -32N\, +3) \Bigr), 
\\
\hspace{-0.95in} && 
 P_{N-2} \, = \, \,
{1 \over 45360} \cdot \, x^{N-2} \cdot \, (x-1)^{N-3}  \cdot \, N (N+1) (N-1) \times 
\nonumber \\
\hspace{-0.95in} && \quad \quad
 \qquad \quad \Bigl((N-2)  \cdot \, (35\,{N}^{5}\, -371\,{N}^{4}\, +1564\,{N}^{3}\, 
-3676\,{N}^{2}\, +4320\,N\, -2448) \cdot \,  {x}^{3}
 \nonumber \\
\hspace{-0.95in} &&  \quad
 \quad  \quad  
 +3\,(N-3) \cdot \,  (N+2) \,  
(35\,{N}^{4}\, -280\,{N}^{3}\, +772\,{N}^{2}\, -929\,N\, +135) \cdot \, {x}^{2}
 \nonumber \\
\hspace{-0.95in} &&  \quad \quad \quad 
+3  \cdot \,  (N+2)  \cdot \, 
(35\,{N}^{5}\, -175\,{N}^{4} \, -194\,{N}^{3}\, +2110\,{N}^{2}\, -2748\,N\, +603) \cdot \,  x
 \nonumber \\
\hspace{-0.95in} && \quad  \quad \quad + \, (N+2) \cdot \,  
\, (35\,{N}^{5}\, +119\,{N}^{4}\, -578\,{N}^{3}\, -1175\,{N}^{2}\, +2682\,N\, -954)
  \Bigr).
 \nonumber
\end{eqnarray}

\section{Non analytical scaling of $ \,  C(N, N)$}
\label{nonanal}

Seeking a logarithmic solution of (\ref{sigmascale2}), one obtains two
solutions that depend on the parameter $e_1$
\begin{eqnarray}
\label{Csacalcons}
\hspace{-0.95in} && \quad  \quad  \quad 
C_{scal} (x)\,  \,  \,=\, \,  \,  \, \,   
{\rm const.} \cdot \, \sum_{k=0}^{\infty}  \, ( \pm 1)^k  \cdot \, S_k (\pm x)
 \cdot \,  \left(\pm {1 \over 4} \,   \ln(x) \,   +e_1 \right)^k, 
\end{eqnarray}
The matching with the first terms given in~\cite{mc-tr-wu-77}, fixes the parameter
$ \,  e_1 = \,   \ln(2) \,  -\gamma/4$.

The first $\, S_k(x)$ read:
\begin{eqnarray}
\label{S0x}
\hspace{-0.95in} &&   
S_0(x) \, \, =\,  \, \,  
1  \, \,  + {1 \over 64} \cdot \,   x^2  \,  +{1 \over 2^{15} } \cdot \,   x^4 \,  
 - {1 \over 2^{17} } \cdot \,   x^5
-{5 \over 2^{21} \cdot 3}\cdot \,  x^6 \, -{1 \over 2^{23}} \cdot \, x^7 \, 
-{469 \over 2^{34} \cdot 3}\cdot \,  x^8
 \,\,   +  \,  \cdots,  
\nonumber \\
\hspace{-0.95in} &&  
S_1(x) \,  \,=\, \,  \,  x \, \,   + {1 \over 64} \cdot \,   x^3 \,  
 +{1 \over 2^{10} }  \cdot\,   x^4 \,   + {5 \over 2^{15} }   \cdot\,  x^5
+ {1 \over 2^{16} } \cdot \,   x^6  \,  + {7 \over 2^{21} \cdot 3 }  \cdot \,  x^7 \, 
  + {35 \over 2^{28} } \,   x^8 
\, \,  +\, \cdots, 
 \nonumber \\
\hspace{-0.95in} && 
S_2(x) \,=\, \, \,   - {1 \over 2^8} \,   x^4  \, \,  -{1 \over 2^{14} }  \cdot\,   x^6 
 \,  - {17 \over 2^{25} } \,   x^8
+ {5 \over 2^{29} } \,   x^9 \, \,  
 - {19 \over 2^{31} \cdot 3 } \cdot \,   x^{10} \, \,  + \cdots,  
 \\
\hspace{-0.95in} && 
 S_3(x) \,=\,  \,  - {1 \over 2^{26}}  \cdot\,   x^9 \, \, 
  -{1 \over 2^{32} }  \cdot\,   x^{11} \, 
  - {37 \over 2^{41} \cdot 3^2 }  \cdot \,  x^{13}
- {13 \over 2^{47} \cdot 3^2 } \cdot \,   x^{15} \,  
 - {13 \over 2^{54} \cdot 3^3 }  \cdot\,   x^{16} 
\, \,  + \,   \cdots,  
\nonumber \\
\hspace{-0.95in} && 
S_4(x) \,=\,  \,   {1 \over 2^{52} \cdot 3^2}  \cdot\,   x^{16}  \, \, 
 +{1 \over 2^{58} \cdot 3^2 } x^{18}\,   + {65 \over 2^{71} \cdot 3^2 }  \cdot\,  x^{20}\,  
+ {67 \over 2^{77} \cdot 3^3 }  \cdot\,  x^{22} \, \,    +\,   \cdots, 
\nonumber \\
\hspace{-0.95in} &&  
S_5(x) \,=\, \,   
{1 \over 2^{90} \cdot 3^4}  \cdot\,  x^{25}\, \, 
  +{1 \over 2^{96}\cdot 3^4 } \cdot \,   x^{27}\,   
+ {101 \over 2^{105} \cdot 3^4 \cdot 5^2 } \cdot \,   x^{29}\,  
+ {103 \over 2^{111} \cdot 3^5 \cdot 5^2 } \cdot \,   x^{31}\, \,    + \, \, \cdots
 \nonumber
\end{eqnarray}

The series $\,  S_k(x)$ begin as
$\,  S_k(x)\,= \,\,  A_{k} \cdot \, x^{k^2} \,  + \,  \cdots$ 
At the order $\,  k^2\,  +2k$, both the even and
the odd orders occur. In between $\,  x^{k^2}$ and $\,  x^{k^2+2k}$, only the coefficients
of $\,  x^{k^2+2p}$ occur (exception of $\,  S_0$). 
This scheme yields that  $\, S_k(x)$ writes as (with $k \ge 1$):
\begin{eqnarray}
\label{Skx}
\hspace{-0.95in} && \quad  \quad  \quad  \quad 
S_k (x)\,\,  \,=\,\, \,\,   A_k \cdot \,  x^{k^2} \cdot \, 
 \Bigl(1 \,  + \, \sum_{p=1}^k a^{(k)}_{2 p} \cdot \,  x^{2\,p} +
  \sum_{p=2 k +1}^\infty b^{(k)}_p \cdot x^p  \Bigr) 
\end{eqnarray}

From the first small series of $S_k(x)$ that we have produced, we infer
the following coefficients:
\begin{eqnarray}
\label{Ak}
\hspace{-0.95in} && \, \quad
{\frac{1}{A_k}} \, \, = \, \,\,  \, 
  (-1)^{k(k+3)/2} \cdot \, 2^{4k(k-1)} \cdot \, \prod_{j=1}^k \, \Gamma(j)^2 , \\
\hspace{-0.95in} && \, \quad 
a^{(k)}_2 \, =\,\,   {\frac{1}{64}}, 
\qquad \qquad \quad 
a^{(k)}_4 \, \, = \,\, \,   {\frac{1\,  +4k^2}{2^{15} \cdot k^2}},
 \quad \quad \, \, \, k >\,   1, 
 \nonumber  \\
\hspace{-0.95in} && \, \quad 
a^{(k)}_6 \, \, = \, \,\,   {\frac{3\,  +4k^2}{2^{21} \cdot 3 \cdot k^2}},
 \quad k \ge\,   1, 
\qquad  
a^{(k)}_8 \, = \,\,   
{\frac{51\,  -58 k^2\,  -16 k^4\,  +32 k^6}{2^{32} \cdot \,  3 \cdot \, k^2 \, (k^2-1)^2}}, 
\quad k \ge \, 2. 
 \nonumber 
\end{eqnarray}

\section{The next-to-diagonal $\, C^{(j)}(N,\, N+1)$ Ising form factors}
\label{AppCjNNp1}

The form factors $\, C^{(j)}(N,M)$ for the anisotropic lattice, are
given, in~\cite{Orrick}. For the isotropic case the result is
\begin{eqnarray}
\label{involved}
\hspace{-0.95in} && \quad  \qquad 
C^{(j)}(M, \, N) \,  \, = \, \, \, \,
 \, {1 \over {j!}} \, 
\int_{-\pi}^{\pi}\, {{d\phi_1} \over {2 \, \pi}} \, \cdots \, 
\int_{-\pi}^{\pi}\, {{d\phi_j} \over {2 \, \pi}} \, 
\Bigl(  \prod_{n=1}^{j} \, {{1} \over {\sinh \gamma_n}} \Bigr)
 \nonumber \\
\hspace{-0.95in} && \quad  \qquad  \qquad 
 \qquad \,  \, \times  \, 
\Bigl(  \prod_{1 \le i \le k \le j} \, h_{ik} \Bigr)^2 \,
 \Bigl( \prod_{n=1}^{j} \, x_n \Bigr)^M \, 
\cos\Bigl( N \, \sum_{n=1}^{j} \,\phi_n \Bigr), 
\end{eqnarray}
with :
\begin{eqnarray}
\label{xn}
\hspace{-0.95in} && \quad 
x_n \,\,\, = \, \,\, \, \,
{{1} \over {2 w}} \, \, -\cos \phi_n \, \,
 -\Bigl( ( {{1} \over {2 w}}  -\cos \phi_n )^2 \, -1 \Bigr)^{1/2}, \\
\hspace{-0.95in} && \quad   
 \sinh \gamma_n \, \,= \,\, \,
\Bigl(  ( {{1} \over {2 w}}  -\cos \phi_n )^2 \, -1 \Bigr)^{1/2}, 
\quad \quad  \, \,
  h_{ik}\,\, = \,\,\, {{ 2 \, (x_i\, x_k)^{1/2} \, 
 \sin((\phi_i - \phi_k)/2)  } \over {1 \, -x_i\, x_k }},
\nonumber 
\end{eqnarray}
with $\, w=\, s/2/(1+s^2)$, and where $\, s$ denotes $\, \sinh(2\, K)$.

\subsection{The linear differential equations of 
$\, C^{(j}(N,\, N+1)$,  $\,  \, j=\, 1,  \,2,  \,3$}
\label{CjNNp1}

We give the linear differential equations that annihilate the first 
next-to-diagonal $\, C^{(j)}(N,\, N+1)$ form factors ($j =\, 1,\,  2, \, 3$).

The first terms of $\, C^{(1)}(N,\, N+1)$ read (with $\, x=\, w^2$)
\begin{eqnarray}
\label{C1N}
\hspace{-0.95in} &&   \,  
 \quad C^{(1)}(N,\, N+1) \,  \,  \, =\, \,  \, \, \, 
 {\frac{2\Gamma(2+2N)}{\Gamma(1+N)\Gamma(2+N)}} \cdot \,x^{N+1} 
\cdot  \, \Bigl(1\,  + {\frac{2 \,(3+2N)^2}{2+N}} \cdot \, x \,
 \nonumber \\
\hspace{-0.95in} && \quad  \quad 
 + {\frac{4 \,(3+2N) \,(5+2N)^2}{3+N}} \cdot   \, x^2 \,  \, \,
+{\frac{8 \,(3+2N) \, (5+2N)^2\, (7+2N)^2}{3 \,(2+N)\, (4+N)}} \cdot  \, x^3 \Bigr).
\end{eqnarray}

These series are annihilated by an order-three  ODE whose corresponding
linear differential operator reads for generic $N$ 
(and written in the variable $ \, s$, where $D_s$ is the derivative $d/ds$)
\begin{eqnarray}
\label{V3}
\hspace{-0.95in} && \quad 
\qquad {\cal V}_3 =\, \, 
 V_2 \cdot V_1, \qquad \qquad \quad V_1 \,=\, D_s, \\
\hspace{-0.95in} && \quad 
 \qquad V_2 \, \, =\,\, \,  \,  \,  
D_s^2\, \, \,  \, + {\frac{1-5s^4}{s (1-s^4)}} \cdot \,  D_s \, \,  \, 
+ \, {\frac{3s^6\, -7s^4\, -3s^2\, -1}{s^2 \cdot \, (1-s^4)^2}}
\, \,  \,  - {\frac{4\,  N\,  (N+1)}{s^2}}.
 \nonumber 
\end{eqnarray}

The form factors $\, C^{(2)}(N,\, N+1)$ expands as (with $ \,x= \, w^2$)
\begin{eqnarray}
\label{C2NN}
\hspace{-0.97in} &&  
  C^{(2)}(N,\, N+1) \, \, \, = \,\,  \,   \, \, \, 
{x}^{ 2 N+3} \cdot \,    {\frac {2 \, (2+N)^{3} \, 
(\Gamma  (2\,N+3))^{2}}{(\Gamma (3+N))^{4}}}  \, \times 
 \\
\hspace{-0.97in} &&  
 \quad 
\Bigl(1\,  \,
+{\frac {2 \, (2\,N+3)  \cdot \, (2\,N+5)^{2}}{ (2+N)\, (3+N) }} \cdot  \,x  
\, \, +{\frac {4 \, (16\,{N}^{3}\,+148\,{N}^{2}\,+456\,N\,+477) \, (2\,N+3) }{
 \, (3+N)  \, (4\,+N) }}\cdot  \, x^2
 \nonumber \\
\hspace{-0.97in} &&  
 \quad +{\frac {8 \, (2\,N+7)^{2} \,
\, (16\,{N}^{4}\,+204\,{N}^{3}\, +956\,{N}^{2}\, +1983\,N\,+1521)\,
  \, (2\,N+3) \, \, (2\,N+5) }
{3 \, (3+N) ^{2} \, (2+N)  \, (4+N)  \, (5+N) }} \cdot  \,x^3\Bigr), 
 \nonumber
\end{eqnarray}
and are annihilated by an order six linear differential operator
 whose corresponding
differential operator reads 
\begin{eqnarray}
\label{V6}
\hspace{-0.95in} && \quad \qquad \qquad \qquad 
{\cal V}_6 \, \, \,=\, \, \, \,  V_3 \cdot  \,V_2 \cdot \, V_1, 
\end{eqnarray}
where:
\begin{eqnarray}
\label{V3bis}
\hspace{-0.95in} && \, \quad  \quad 
 \, V_3 \,\, = \,\,\, \, \, D_s^3\,\,\, 
\, + {\frac{4(1-5s^4)}{s (1-s^4)}} \cdot \, D_s^2\, 
 \\
\hspace{-0.95in} && 
 \quad \quad  \quad \quad 
+ \, \Bigl({\frac{105s^8\, -16s^6\, -178s^4\, -16s^2\, -7}{
s^2 \cdot \,  (1\, -s^4)^2}} 
\, -{\frac{16 N (N+1)}{s^2}}\Bigr) \cdot \,  D_s 
\nonumber \\
\hspace{-0.95in} &&
 \quad \quad  \quad \quad 
- \, \Bigl(3 \, {\frac{45s^{12}\, -32s^{10}\, -199s^8\, 
-96s^6\, +87s^4\, +3}{s^3 \cdot \, (1-s^4)^3}}\, 
\,  + {\frac{48\,  N \,  (N+1)}{s^3}}\Bigr). 
\nonumber
\end{eqnarray}

The first terms of $\, C^{(3)}(N,\, N+1)$ read (with $\, x=\, w^2$)
\begin{eqnarray}
\label{C3NN}
\hspace{-0.95in} &&   \quad \quad
C^{(3)}(N,\, N+1) \, \, \, =\, \,  \,\,\,
3072  \cdot \, {\frac { (3+N)^{2}{64}^{N} \, (\Gamma (N+5/2))^{3} 
\cdot \, {x}^{6+3\,N}}{{\pi }^{3/2}
 \, (3+2\,N)^{2}\cdot  \, (\Gamma (N+4))^{3}}}   \, \times 
\nonumber \\
\hspace{-0.95in} && \quad   \quad \quad  \qquad  \Bigl(
1\, \, +6\,{\frac {(7+2\,N)^{2} \, (2+N) \cdot \,  x}{(N+4) \,  (3+N) }} \,
 \\
\hspace{-0.95in} && \quad   \quad \quad  \qquad 
 +36\,{\frac { (5+2\,N) 
 \, (4\,{N}^{4}\, +56\,{N}^{3}\, +287\,{N}^{2} \, 
+636\,N \, +507)\cdot  \, {x}^{2}}{(3+N)  \, (N+4) \, 
 \, (N+5) }}   \Bigr),
  \nonumber
\end{eqnarray}
and are solution of an order-ten  ODE whose corresponding
linear differential operator factorizes as
\begin{eqnarray}
\label{V10}
\hspace{-0.95in} && \quad  \qquad \qquad 
{\cal V}_{10} \, \, =\, \,\, \,\, 
V_4 \cdot \, V_3 \cdot  \, V_2 \cdot  \, V_1, 
\end{eqnarray}
with
\begin{eqnarray}
\label{V4}
\hspace{-0.95in} && \quad  \qquad \qquad 
V_4  \,\, =\, \,\,\,\,  D_s^4 \,\, \, + {\frac{p_3}{p_4}} \cdot \, D_s^3 \,\, 
 +\,{\frac{p_2}{p_4}} \cdot \, D_s^2\, \, 
 +\, {\frac{p_1}{p_4}} \cdot \, D_s\, \, 
+\,{\frac{p_0}{p_4}}, 
\end{eqnarray}
where:
\begin{eqnarray}
\label{p4}
\hspace{-0.95in} && \quad    
p_4 \,=\,\,\,  s^4 \cdot \, (1\, -s^4)^3 \cdot \, (1\, +s^2),
 \nonumber \\
\hspace{-0.95in} && \quad  
p_3 \,=\,\, \,  10 \cdot \, s^3 \cdot \, (1\, -s^4)^2 \cdot \,(1\, +s^2) \cdot \, (1\, -5 s^4), 
\nonumber \\
\hspace{-0.95in} && \quad  
p_2 \,=\,\, \,
-s^2 \cdot \, (1\, -s^4) \cdot \, (1\, +s^2) \cdot \, 
 \Bigl(40\,\,  (1\, -s^4)^2 \cdot \,  N \cdot \, (N+1)\,\,\, +17\,+40 s^2
\nonumber \\
\hspace{-0.95in} && \quad \quad   \qquad  \qquad  \,+998 s^4 
+40 s^6-823 s^8 \Bigr),
 \\
\hspace{-0.95in} && \quad  
p_1 \,=\,\,\,  s \cdot \, (1+s^2)\cdot \, 
 \Bigl(-8 \,(1-s^4)^2 (47-83 s^4) \cdot \,  N \cdot \, (N+1)\, 
  -175 -72 s^2
\nonumber \\
\hspace{-0.95in} && \quad  \qquad  \qquad  \quad 
-3243 s^4 +2112 s^6\, +16803 s^8\, +968 s^{10}\, -5193 s^{12} \Bigr), 
\nonumber \\
\hspace{-0.95in} && \quad  
p_0 \,=\, 
144 \cdot \, (1-s^4)^3  \,(1+s^2) \cdot \, N^4\, \, 
+288 \, (1-s^4)^3 \,(1+s^2) \cdot \,  N^3
 \nonumber \\
\hspace{-0.95in} && \quad  \quad \quad \quad  
-144 \cdot \,(1-s^4) (1+s^2)\cdot \,  
\Bigl(5 \, -2 s^2 \, -50 s^4 \, -2 s^6 \,+17 s^8 \Bigr) \cdot \,   N^2 
\nonumber \\
\hspace{-0.95in} && \quad \quad   \quad \quad  
-288 \cdot \, (1-s^4) (1+s^2)\cdot \,  
\Bigl(3 \,-s^2 \,-26 s^4 \,-s^6 \,+9 s^8 \Bigr) \cdot \,  N
 \nonumber \\
\hspace{-0.95in} && \quad  \quad \quad \quad  
+48 \, s^2 \cdot \,   
\Bigl( 6\, +105 s^2\, +63 s^4\, +1705 s^6\, +1247 s^8 \, -110 s^{10}\, -216 s^{12} \Bigr).
 \nonumber
\end{eqnarray}

{\bf Remark:} Unlike what we have seen for the diagonal 
$\, f^{(j)}_N = C^{(j)}(N,\, N)$, one notes that the linear
differential equation $\, {\cal V}_6$, which annihilates $\, C^{(2)}(N,\, N+1)$, solves
$\, C^{(1)}(N,\, N+1)$ as well. Also, the  linear differential 
equation $\, {\cal V}_{10}$ which annihilates $\, C^{(1)}(N,\, N+1)$ 
and $ \, C^{(3)}(N,\, N+1)$, solves $\, C^{(2)}(N,\, N+1)$ as well.

\subsection{The linear differential equations in the scaling limit}
\label{lindiffinscal}

The scaling limit is obtained by performing the variable change $\, s=\, 1\, -y/N$,
keeping the leading terms in $\, N$.
However, since for the diagonal form factors the variable change was 
$\, t =\,  1\, -x/N$ and since $\, t =\,  s^4$, we will take, 
for easy comparison, the following variable
change $\,\,  s =\,\,   1\, -x/(4N)$.

The scaling limit of $\, {\cal V}_{3}$ (corresponding to $\, C^{(1)}(N,\,  N+1)$) 
has a {\em direct sum} factorization:
\begin{eqnarray}
\label{V2V1}
\hspace{-0.95in} && \quad \quad  \qquad   \qquad  
(V_2 \cdot V_1)^{scal}  \, \,  \, = \,\, \, \,   \,   L_1^{scal} \oplus \,  L_2^{scal} 
\end{eqnarray}
Note the linear differential operators at the right-hand-side which are the operators 
given in \ref{Lnscal}. $\, L_2^{scal}$ is the scaling limit of the operators of 
the diagonal $\, f^{(1)}_N$.

The scaling limit of $\, {\cal V}_{6}$ (corresponding to
 $\, C^{(2)}(N, N+1)$) has {\em also a direct sum decomposition}:
\begin{eqnarray}
\label{V3V2V1}
\hspace{-0.95in} && \quad  \qquad  \qquad   
(V_3 \cdot V_2 \cdot V_1)^{scal}  \, \,   \,\, = \,\,\, \,  \,  \,  \, 
  L_1^{scal} \oplus \, L_2^{scal} \oplus\, \tilde{L}_3^{scal}. 
\end{eqnarray}
Here also, $\, \tilde{L}_3^{scal}$ is the operator appearing 
in \ref{Lnscal} and corresponding to the
scaling limit of the operator for $\, f^{(2)}_N$.

The scaling limit of $\, {\cal V}_{10}$ (which corresponds to 
$\, C^{(1)}(N, N+1)$, $\, C^{(2)}(N, N+1)$, and $\, C^{(3)}(N, N+1)$) 
factorizes as:
\begin{eqnarray}
\label{scalV4321}
\hspace{-0.95in} && \,  \qquad   \qquad   \,  \, 
(V_4 \cdot \, V_3 \cdot\,  V_2 \cdot \, V_1)^{scal} 
 \, \,  \,  \, =\,\, \,  \,  \,  \, 
 L_1^{scal} \oplus \, \tilde{L}_3^{scal} \oplus \, L_4^{scal} \cdot \, L_2^{scal}.
\end{eqnarray}
Again, in the scaling limit, the linear differential operator corresponding 
to the diagonal $\, f^{(3)}_N$ appears.

\section{Rational solutions of the symmetric (or exterior) 
square of $\, L^{scal}_n$}
\label{ratioscal}

For $\, n$ odd, the symmetric square of $\, L^{scal}_n$
annihilate the following rational solutions $\, r_n^{scal}$
\begin{eqnarray}
\label{solrat}
\hspace{-0.95in}  \, 
&&r_3^{scal} \, =\,  \, \, {\frac {{x}^{2}+1}{{x}^{2}}},
 \quad \quad \quad \quad 
r_5^{scal} \, =\,  \, \, {\frac {{x}^{6}+4\,{x}^{4}-17\,{x}^{2}-9}{{x}^{6}}},  
\nonumber \\
\hspace{-0.95in}  \, 
&&r_7^{scal} \, =\,  \, \, 
{\frac {{x}^{14}+9\,{x}^{12}-105\,{x}^{10}+1122\,{x}^{8}
+2754\,{x}^{6}-15822\,{x}^{4}-4347\,{x}^{2}+2916}{{x}^{14}}},
\nonumber \\
\hspace{-0.95in}  \, 
&&x^{24} \cdot r_9^{scal} \, =\,  \, \,\, 
{x}^{24}\, +16\,{x}^{22}\, -354\,{x}^{20}\, +9486\,{x}^{18}\,
 -158364\,{x}^{16}\, -1230840\,{x}^{14} 
\nonumber \\
\hspace{-0.95in} && \, \quad \quad 
+20545650\,{x}^{12}\,-170513100\,{x}^{10}\,
 -305967375\,{x}^{8}\, +552217500\,{x}^{6}
\nonumber \\
\hspace{-0.95in} && \,\quad \quad 
-1474858125\,{x}^{4}\, -603703125\,{x}^{2}\, +218700000. 
\end{eqnarray}

For $\, n$ even, the exterior square of $\, L^{scal}_n$
annihilate the following rational solutions $\, r_n^{scal}$
\begin{eqnarray}
\label{solrat2}
\hspace{-0.95in}  \, 
&&r_4^{scal} \, =\,  \, \, 
{\frac {{x}^{2}-2}{{x}^{3}}},  \quad \quad \quad \quad 
r_6^{scal} \, =\,  \, \, 
{\frac {4\,{x}^{8} \,-24\,{x}^{6}\, +240\,{x}^{4}\, +144\,{x}^{2}\, +243}{{x}^{9}}}, 
\nonumber \\
\hspace{-0.95in}  \, 
&&x^{17} \cdot r_8^{scal} \, = \,  \, \,\,  
{x}^{16}\, -12\,{x}^{14}\, +300\,{x}^{12}\, -5220\,{x}^{10}\, -13734\,{x}^{8}\,
-15804\,{x}^{6}\, +482760\,{x}^{4} 
\nonumber \\
\hspace{-0.95in} && \, 
+90720\,{x}^{2}\, +327240,
\nonumber \\
\hspace{-0.95in}  \, 
&&x^{29} \cdot r_{10}^{scal} \, =\,  \, \, \, 
8\,{x}^{28}\, -160\,{x}^{26}\, +7200\,{x}^{24} \,-298080\,{x}^{22}\,
+7083540\,{x}^{20}\, +52569000\,{x}^{18}
 \nonumber \\
\hspace{-0.95in}  \, && \quad \quad 
-31590000\,{x}^{16}\, -7017678000\,{x}^{14} \,+113926753125\,{x}^{12}
+82387935000\,{x}^{10}
\nonumber \\
\hspace{-0.95in}  \, &&\quad \quad 
+552374235000\,{x}^{8}\, -4206913200000\,{x}^{6}\, -1635103715625\,{x}^{4}
\nonumber \\
\hspace{-0.95in}  \, &&\quad \quad 
-1234903218750\,{x}^{2} \,+3690562500000.
\end{eqnarray}

For $\, n$ odd, the symmetric square of the adjoint 
of the $\, L^{scal}_n$
annihilate the following  {\em polynomial} solutions $\, R_n^{scal}$
\begin{eqnarray}
\label{solratadj}
\hspace{-0.95in}  \, \,
R_3^{scal} \, &=&\,  \, \, {\frac {1}{{x}^{2}}}, 
\quad \quad \quad \quad  \quad 
R_5^{scal} \, =\,  \, \, {x}^{6} \,-12\,{x}^{4}\, -18\,{x}^{2}\, -18, \quad 
\nonumber \\
\hspace{-0.95in}  \, \,
R_7^{scal} \, &=&\,  \, \, 
{x}^{4} \cdot \, (2\,{x}^{18} -360\,{x}^{16} +15390\,{x}^{14} -171450\,{x}^{12}
+251100\,{x}^{10}-437400\,{x}^{8} \nonumber \\
\hspace{-0.95in} && \, 
-2673000\,{x}^{6} -7290000\,{x}^{4} -9021375\,{x}^{2} -2916000), 
\nonumber \\
\hspace{-0.95in}  \, \,
R_9^{scal} \, &=&\,  \, \, 
{x}^{10} \cdot \, (4\,{x}^{36} -3360\,{x}^{34} +1010800\,{x}^{32} -141699600\,{x}^{30}
+10025694000\,{x}^{28} \nonumber \\
\hspace{-0.95in}  \,\, &&
-361966096800\,{x}^{26}+6415388406000\,{x}^{24} -50642114166000\,{x}^{22} 
 \nonumber \\
\hspace{-0.95in}  \,\, &&
+144901264095000\,{x}^{20} -65747358180000\,{x}^{18} +157049519760000\,{x}^{16} 
\nonumber \\
\hspace{-0.95in}   \,\, &&
+1294768370700000\,{x}^{14} +9174749528700000\,{x}^{12} 
+49115481975000000\,{x}^{10}
 \nonumber \\
\hspace{-0.95in}  \,\, &&
+178940697027750000\,{x}^{8} +406258455983625000\,{x}^{6} 
+504390999090234375\,{x}^{4}
\nonumber \\
\hspace{-0.95in}  \,\, &&
+271021254918750000\,{x}^{2} +36456852600000000).
\end{eqnarray}

For $\, n$ even, the exterior square of the {\em adjoint} 
of the $\, L^{scal}_n$
annihilate the following  {\em polynomial} solutions $\, R_n^{scal}$
\begin{eqnarray}
\label{solratadj2}
\hspace{-0.95in}   
&&R_4^{scal} \, =\,  \, \,
 x \cdot \, ( {x}^{2}-2), 
 \quad
R_6^{scal} \, =\,  \, \, 
{x}^{5} \cdot \, 
(x^{10} \,-72\,{x}^{8}\, +792\,{x}^{6}\, -720\,{x}^{4}\, -1377\,{x}^{2}\, -1134), 
\nonumber \\
\hspace{-0.95in} 
&&R_8^{scal} \, =\,  \, \, {x}^{11} \cdot \, 
( 4\,{x}^{24} \,-1800\,{x}^{22} \,+259200\,{x}^{20} \,-15170400\,{x}^{18}\, +370747125\,{x}^{16}
 \nonumber \\
\hspace{-0.95in} && \,\quad \quad  \quad 
-3461802300\,{x}^{14}\, +8998897500\,{x}^{12}\,
 -1729552500\,{x}^{10} \,+6142736250\,{x}^{8}
 \nonumber \\
\hspace{-0.95in} && \,\quad \quad \quad 
+36349762500\,{x}^{6} \,+83751165000\,{x}^{4}\,
 +82668600000\,{x}^{2}\, +22471425000),
\nonumber \\
\hspace{-0.95in} 
&&R_{10}^{scal} \, = \,  \, \, {x}^{19} \cdot \, (4\,{x}^{44} -6272\,{x}^{42}
+3857280\,{x}^{40} -1226991360\,{x}^{38} +223629630000\,{x}^{36} 
\nonumber \\
\hspace{-0.95in} && \, \quad \quad  \quad 
-24414033420000\,{x}^{34}\,+1620296469590400\,{x}^{32}\,
 -64985230800000000\,{x}^{30}
 \nonumber \\
\hspace{-0.95in} && \, \quad \quad  \quad 
+1535802293434972500\,{x}^{28} \,-20371435267457610000\,{x}^{26} 
\nonumber \\
\hspace{-0.95in} && \, \quad \quad  \quad 
+139451099666404050000\,{x}^{24}\, -424167698945936610000\,{x}^{22} 
\nonumber \\
\hspace{-0.95in} && \, \quad \quad  \quad 
+409086586150282687500\,{x}^{20} \,-24200532938759625000\,{x}^{18}  
\nonumber \\
\hspace{-0.95in} && \, \quad \quad  \quad 
+246154709199372750000\,{x}^{16}\, +1920553025539034250000\,{x}^{14} 
 \nonumber \\
\hspace{-0.95in} && \, \quad \quad  \quad 
+11369086319287068750000\,{x}^{12}\, +49298397042819015000000\,{x}^{10}  
\nonumber \\
\hspace{-0.95in} && \, \quad \quad  \quad 
+143327291046534157500000\,{x}^{8}\, +257142668429453821875000\,{x}^{6}  
\nonumber \\
\hspace{-0.95in} && \, \quad \quad \quad 
+252273151762962774609375\,{x}^{4} \,+109262436886268613281250\,{x}^{2} 
 \nonumber \\
\hspace{-0.95in} && \, \quad \quad  \quad 
+12887098647274687500000).
\end{eqnarray}

\end{document}